\documentclass[a4paper,11pt]{article}

\usepackage{amsmath,amsthm,amsbsy,amssymb,amsfonts,graphicx,mathrsfs,bm,cite,color}
\usepackage{hyperref}
\usepackage[utf8]{inputenc}

\makeatletter
\catcode`\@=11
\@addtoreset{equation}{section}

\makeatother

\makeatletter

\@addtoreset{table}{section}
\makeatother

\renewcommand{\thefootnote}{\arabic{footnote}}

\newcommand{\Exp}[1]{\operatorname{e}^{#1}}
\newcommand{\abs}[1]{\lvert {#1} \rvert}
\newcommand{\rmd}{{\mathrm{d}}}
\newcommand{\nn}{\nonumber}
\newcommand{\Lie}{\pounds}
\newcommand{\gLie}{\hat{\pounds}}

\newcommand{\cC}{\mathcal C}
\newcommand{\cE}{\mathcal E}\newcommand{\cF}{\mathcal F}
\newcommand{\cH}{\mathcal H}
\newcommand{\cI}{\mathcal I}\newcommand{\cJ}{\mathcal J}
\newcommand{\cK}{\mathcal K}
\newcommand{\cM}{\mathcal M}
\newcommand{\cP}{\mathcal P}

\newcommand{\WSa}{\alpha}
\newcommand{\WSb}{\beta}
\newcommand{\WSc}{\gamma}

\newcommand{\MA}{C}

\newcommand{\SL}{\text{SL}}
\newcommand{\GL}{\text{GL}}
\newcommand{\OO}{\text{O}}

\allowdisplaybreaks[3]

\begin{document}

\begin{titlepage}
\renewcommand{\thefootnote}{\fnsymbol{footnote}}

\vspace*{1.0cm}

\begin{center}
\textbf{\Large Non-Abelian $U$-duality for membrane}%
\end{center}
\vspace{1.0cm}

\centerline{\large
{Yuho Sakatani}%
\footnote{E-mail address: \texttt{yuho@koto.kpu-m.ac.jp}}
\ \ and \ \ 
{Shozo Uehara}%
\footnote{E-mail address: \texttt{uehara@koto.kpu-m.ac.jp}}
}

\vspace{0.2cm}

\begin{center}
{\it Department of Physics, Kyoto Prefectural University of Medicine,}\\
{\it Kyoto 606-0823, Japan}
\end{center}

\vspace*{2mm}

\begin{abstract}
$T$-duality of string theory can be extended to the Poisson--Lie $T$-duality when the target space has a generalized isometry group given by a Drinfel'd double. In M-theory, $T$-duality is understood as a subgroup of $U$-duality, but the non-Abelian extension of $U$-duality is still a mystery. In this paper, we study membrane theory on a curved background with a generalized isometry group given by the $\cE_n$ algebra. This provides a natural setup to study non-Abelian $U$-duality because the $\cE_n$ algebra has been proposed as a $U$-duality extension of the Drinfel'd double. We show that the standard treatment of Abelian $U$-duality can be extended to the non-Abelian setup. However, a famous issue in Abelian $U$-duality still exists in the non-Abelian extension.
\end{abstract}

\thispagestyle{empty}
\end{titlepage}

\setcounter{footnote}{0}

\newpage

\tableofcontents

\newpage

\section{Introduction}
\label{sec:intro}

Abelian $T$-duality is a symmetry of string theory when the target space has $D$ commuting Killing vector fields. 
This $T$-duality can be extended to the Poisson--Lie (PL) $T$-duality \cite{hep-th:9502122,hep-th:9509095} when the target geometry has a certain symmetry generated by the Lie algebra of the Drinfel'd double. 
For the PL $T$-duality, the usual Killing vector fields are not necessary and we can consider the extended $T$-duality in a more general class of target spaces. 
Similar to Abelian $T$-duality, the PL $T$-duality is a symmetry of the supergravity equations of motion (see e.g.~\cite{hep-th:0106211}), and it can generate various supergravity solutions (see e.g.~\cite{hep-th:9509095,hep-th:9803175,hep-th:9903152,hep-th:0205245,hep-th:0210095,hep-th:0403164,hep-th:0408126,hep-th:0601172,hep-th:0608069,1201.5939,1308.0153,1903.12175,1905.13627,1910.08436}). 
To be more precise, some dual geometries do not solve the standard supergravity equations but rather the modified ones, known as the generalized supergravity equations \cite{1511.05795,1605.04884}.\footnote{Note that the modified equations for the NS--NS sector have been discussed earlier in \cite{Hull:1985rc,hep-th:9409011}.} 
However, as shown in \cite{1611.05856,1703.09213}, the generalized supergravity equations can be derived from double field theory (DFT) \cite{hep-th:9302036,hep-th:9305073,hep-th:9308133,0904.4664,1006.4823}, which is a $T$-duality-manifest formulation of supergravity. 
Thus, now the PL $T$-duality has been understood as a symmetry of DFT \cite{1707.08624,1708.04079,1810.07763,1810.11446,1903.12175}. 
Recently, various aspects of the PL $T$-duality, in particular, its relation to the Yang--Baxter deformation \cite{hep-th:0210095,0802.3518,1308.3581,1309.5850,1401.4855} (i.e.~a class of integrable deformations of string theory) have been clarified and still actively studied. 

Type IIA string theory compactified on a flat $D$-torus has the $\OO(D,D)$ Abelian $T$-duality symmetry. 
From the perspective of M-theory compactified on a flat $n$-torus ($n\equiv D+1$), this $\OO(D,D)$ Abelian $T$-duality group is understood as a subgroup of the $E_n$ $U$-duality group. 
By using the $U$-duality-manifest formulation of supergravity, known as the exceptional field theory (EFT) \cite{1206.7045,1308.1673,1312.0614,1312.4542,1406.3348}, the $U$-duality symmetry in supergravity has been clearly understood. 
EFT also exhibits the duality between M-theory and type IIB supergravity, and it also provides a useful framework to study various non-geometric backgrounds or non-trivial compactifications. 
Moreover, by using EFT, a $U$-duality extension of the PL $T$-duality has been discussed recently in \cite{1911.06320,1911.07833} for $n\leq 4$, where the Drinfel'd double is realized as a subalgebra of the proposed $\cE_n$ algebra. 
There still remain many things to be clarified, but it has been expected that this $\cE_n$ algebra is the symmetry underlying the non-Abelian extension of $U$-duality. 

In contrast to the success in supergravity, $U$-duality symmetry in membrane theory remains to be mysterious. 
In the case of string theory, the equations of motion in a flat space have been successfully expressed in a $T$-duality-covariant form \cite{Duff:1989tf}. 
By closely following this approach, the equations of motion for a membrane have been expressed in a $U$-duality-covariant form in \cite{Duff:1990hn} (see also \cite{1008.1763,1208.1232}). 
However, as pointed out in \cite{1509.02915}, certain integrability is broken under general $U$-duality transformations, and it has been concluded that only a subgroup of Abelian $U$-duality is the symmetry of the membrane equations of motion. 
Only when the dimension of the target space is $n=3$ (where the membrane is spacefilling and called the topological membrane), the full $\SL(2)\times\SL(3)$ $U$-duality is consistently realized \cite{1509.02915} (see also \cite{hep-th:9407021,hep-th:9512203,hep-th:9704178}).

In this paper, focusing on the successful case $n=3$, we investigate non-Abelian $U$-duality in membrane theory. 
Our main results are as follows. 
In the $T$-duality-covariant formulation of string theory, the displacement $\rmd x^m(\sigma)$ $(m=1,\dotsc,D)$ is extended to the generalized displacement $\cP^M(\sigma)=\rmd x^M(\sigma)$ ($M=1,\dotsc,2D$). 
In the setup of the PL $T$-duality, this $\cP^M$ is further extended to the Maurer--Cartan (MC) form satisfying $\rmd\cP^A = \frac{1}{2}\,\cF_{BC}{}^A\,\cP^B\wedge \cP^C$ ($A=1,\dotsc,2D$), where $\cF_{BC}{}^A$ denote the structure constants of the Drinfel'd double. 
Locally, it can be parameterized as $\cP^A\,T_A = \rmd l\,l^{-1}$ by using a group element of a Drinfel'd double $l(\sigma)$, and this reduces to $\cP^M=\rmd x^M$ when the Drinfel'd double is Abelian. 
In the $U$-duality-covariant formulation of the topological membrane in flat space ($n=3$), the generalized displacement satisfies $\rmd\cP^I=0$ ($I=1,\dotsc,\frac{n(n+1)}{2}$) and it can be locally express as $\cP^I = \rmd x^I$\,. 
This paper studies its extension to the case where the target space is curved. 
By requiring the target space to have a symmetry of the $\cE_n$ algebra, we show that the generalized displacement satisfies the MC equation $\rmd\cP^A = \frac{1}{2}\,\cF_{BC}{}^A\,\cP^B\wedge \cP^C$ ($A=1,\dotsc,\frac{n(n+1)}{2}$), where $\cF_{BC}{}^A$ are the structure constants of the $\cE_n$ algebra. 
The MC equation does not depend on the choice of the generators $T_A$ in the $\cE_n$ algebra, and it is manifestly covariant under the change of generators $T'_A = C_A{}^B\,T_B$\,, where the constant matrix $C_A{}^B$ is an element of the $U$-duality group $E_3\equiv \SL(2)\times\SL(3)$\,. 
This arbitrariness in the choice of generators is what we call non-Abelian $U$-duality. 
It naturally unifies the PL $T$-duality and Abelian $U$-duality. 

For clarity, we here note several subtleties. 
In string theory, the number of the equations of motion for the scalar fields $x^m$ is $D$, and the number of non-trivial MC equation is also $D$. 
Accordingly, we can show that the equations of motion are equivalent to the MC equation. 
In contrast, in membrane theory, the number of the equations of motion for the scalar fields $x^i$ is $n$ while the number of non-trivial MC equation is $\frac{n(n-1)}{2}$ (see section \ref{sec:Abelian-U}). 
For $n\geq 4$, the number of non-trivial MC equation is greater than that of the equations of motion, and it is impossible to realize the full MC equation even under the equations of motion. 
Therefore, we can show the MC equation only when $n=3$\,. 
In $n=3$, both the equations of motion and the MC equation are identically satisfied and it is not so clear whether we can claim that the non-Abelian $U$-duality is the symmetry of the membrane equations of motion. 
However, the situation is completely the same as Abelian $U$-duality. 
Our result is a natural non-Abelian extension of the standard Abelian $U$-duality, and the fact that the generalized displacement $\cP^A$ satisfies the MC equation is non-trivial. 

The structure of this paper is as follows. 
In section \ref{sec:Abelian}, we review Abelian $T$-duality and $U$-duality. 
The famous issue in $U$-duality is also reviewed. 
In section \ref{sec:non-Abelian}, we discuss the non-Abelian extensions. 
The PL $T$-duality in string theory is reviewed in section \ref{sec:PL-T}. 
In section \ref{sec:NAUD}, we discuss non-Abelian $U$-duality in membrane theory. 
Section \ref{sec:discussion} is devoted to the summary and discussion. 
Technical details are given in Appendices.

\section{Abelian $T$-duality and $U$-duality}
\label{sec:Abelian}

In this section, we review Abelian $T$-duality in string theory. 
We also review Abelian $U$-duality in membrane theory and explain a notorious issue specific to $U$-duality. 

\subsection{Abelian $T$-duality in string theory}

In order to perform Abelian $T$-duality, the target space needs to have Abelian Killing vectors. 
Thus, we here consider string theory in a $D$-dimensional flat space, where the supergravity fields are constant. 
The string equations of motion can be expressed as
\begin{align}
 \rmd \cJ_m = 0\,,\qquad \cJ_m \equiv g_{mn} *\rmd x^n + B_{mn}\, \rmd x^n\,.
\end{align}
We also have a trivially conserved current,
\begin{align}
 \rmd \cJ^m = 0\,,\qquad \cJ^m \equiv \rmd x^m\,,
\end{align}
and Abelian $T$-duality can be understood as a permutation of the equations of motion and the Bianchi identities \cite{Duff:1989tf},
\begin{align}
 \cJ_M \ \to\ \cJ'_M = C_M{}^N\,\cJ_N\,,\qquad 
 \bigl(\cJ_M\bigr) \equiv \begin{pmatrix} \cJ_m \\ \cJ^m \end{pmatrix}
 = \begin{pmatrix} g_{mn} *\rmd x^n + B_{mn}\, \rmd x^n \\ \rmd x^m \end{pmatrix} .
\label{eq:string-rotation}
\end{align}
If we introduce the generalized metric
\begin{align}
 \bigl(\cH_M{}^N\bigr) \equiv \begin{pmatrix} (B\,g^{-1})_m{}^n & (g - B\,g^{-1}\,B)_{mn} \\ g^{mn} & -(g^{-1}\,B)^m{}_n \end{pmatrix},
\label{eq:cHdU}
\end{align}
which is an $\OO(D,D)$ matrix preserving the $\OO(D,D)$ metric invariant
\begin{align}
 \cH_M{}^P\,\cH_N{}^Q\,\eta_{PQ} = \eta_{MN}\,,\qquad 
 \bigl(\eta_{MN}\bigr) \equiv \begin{pmatrix} 0 & \delta_m^n \\ \delta^m_n & 0\end{pmatrix} ,
\end{align}
we find that the 1-form fields $\cJ_M(\sigma)$ satisfy the self-duality relation,
\begin{align}
 \cJ_M = \cH_M{}^N * \cJ_N \,.
\label{eq:self-duality-J-string}
\end{align}
In order to keep this relation under the rotation \eqref{eq:string-rotation}, $\cH_M{}^N$ also should be transformed as
\begin{align}
 \cH'_M{}^N = C_M{}^P\,\cH_P{}^Q\,(C^{-1})_Q{}^N \,.
\end{align}
By requiring that the transformed metric $\cH'_M{}^N$ is still an $\OO(D,D)$ matrix, the matrix $C_M{}^N$ is required to be an $\OO(D,D)$ element. 
This $\OO(D,D)$ symmetry is the standard Abelian $T$-duality. 

Now, let us introduce 1-form fields $\cP^M(\sigma)$ through
\begin{align}
 \cJ_M = \cH_{MN}\,* \cP^N \,,\qquad
 \bigl(\cP^M\bigr) = \begin{pmatrix} \rmd x^m \\ g_{mn} *\rmd x^n + B_{mn}\, \rmd x^n 
\end{pmatrix}.
\label{eq:string-P}
\end{align}
Here and hereafter, we raise or lower the indices $M,N$ by using the matrix $\eta$; e.g.~$\cH_{MN}= \cH_{M}{}^P\,\eta_{PN}$\,. 
Under the equations of motion $\rmd \cJ_M=0$, the 1-form fields $\cP^M$ satisfy
\begin{align}
 \rmd * \cP^M = 0 \,. 
\label{eq:d*P=0}
\end{align}
From the self-duality relation \eqref{eq:self-duality-J-string}, $\cP^M$ also satisfy
\begin{align}
 \cP^M = \cH^M{}_N * \cP^N \,.
\label{eq:self-duality-P-string}
\end{align}
Then, since $\cH^M{}_N$ is constant and invertible, the equations of motion \eqref{eq:d*P=0} are equivalent to
\begin{align}
 \rmd \cP^M = 0 \,.
\end{align}
This shows that, under the equations of motion, we can locally express the 1-form fields as
\begin{align}
 \cP^M(\sigma) = \rmd x^M(\sigma)\,,\qquad \bigl(x^M\bigr) = \begin{pmatrix} x^m \\ \tilde{x}_m \end{pmatrix}.
\end{align}
The scalar fields $x^M(\sigma)$ are interpreted as the embedding functions into a $2D$-dimensional doubled space and $\cP^M=\rmd x^M$ is interpreted as the generalized displacement. 

\subsection{Abelian $U$-duality in membrane theory}
\label{sec:Abelian-U}

In \cite{Duff:1990hn}, the same idea has been applied to membrane theory in a flat space. 
By following \cite{Duff:1990hn,1008.1763}, we consider the dynamics of a membrane in an $n$-dimensional Lorentzian spacetime ($n\leq 4$). 
Similar to the string case, the equations of motion are expressed as
\begin{align}
 \rmd \cJ_i = 0 \,,\qquad \cJ_i \equiv g_{ij} * \rmd x^j - \frac{1}{2}\,\MA_{ijk}\,\rmd x^j \wedge\rmd x^k\,,
\label{eq:eom-Abelian-U}
\end{align}
where $i,j=1,\dotsc,n$ and $*$ is the Hodge star operator associated with the induced metric $h_{\WSa\WSb}\equiv g_{ij}\,\partial_{\WSa}x^i\,\partial_{\WSb}x^j$. 
The trivially conserved current, known as the topological current is defined as $\cJ^{ij} \equiv \rmd x^i\wedge\rmd x^j$ and we consider a combination,
\begin{align}
 \bigl(\cJ_I\bigr) \equiv \begin{pmatrix} \cJ_i \\ \frac{\cJ^{i_1i_2}}{\sqrt{2!}} \end{pmatrix}
 = \begin{pmatrix} g_{ij} * \rmd x^j - \frac{1}{2}\,\MA_{ijk}\,\rmd x^j \wedge\rmd x^k \\ \frac{\rmd x^{i_1}\wedge\rmd x^{i_2}}{\sqrt{2!}} \end{pmatrix} .
\label{eq:cJ-Abelian-U}
\end{align}
Similar to Abelian $T$-duality \eqref{eq:string-rotation}, Abelian $U$-duality can be understood as a permutation of the equations of motion ($\rmd \cJ_i=0$) and the Bianchi identities ($\rmd \cJ^{ij}=0$),
\begin{align}
 \cJ_I \ \rightarrow \ \cJ'_I = C_I{}^J\,\cJ_J \,.
\label{eq:U-dual-J}
\end{align}

\paragraph{Definitions:}
In order to see that the matrix $C_I{}^J$ is restricted to the $E_n$ $U$-duality group, let us make several definitions. 
The generalized metric in EFT,\footnote{The generalized metric $\cH_{IJ}$ has ``effective weight'' 0 while $\cM_{IJ}\equiv \abs{g}^{\frac{1}{9-n}}\cH_{IJ}\in E_n$ has weight 0 \cite{1312.0614}.} which is a $U$-duality-covariant combination of supergravity fields, is defined as
\begin{align}
 \bigl(\cH_{IJ}\bigr) &\equiv \begin{pmatrix}
 g_{ij} + \frac{1}{2}\,\MA_i{}^{k_1k_2}\,\MA_{k_1k_2j} & -\frac{\MA_{i}{}^{j_1j_2}}{\sqrt{2!}} \\ -\frac{\MA^{i_1i_2}{}_j}{\sqrt{2!}} & g^{i_1i_2,\,j_1j_2}
 \end{pmatrix},
\end{align}
where $g^{i_1i_2,\,j_1j_2}\equiv g^{i_1[j_1}g^{j_2]i_2}$ and the inverse is denoted as
\begin{align}
 \bigl(\cH^{IJ}\bigr) &= \begin{pmatrix} g^{ij} & \frac{\MA^i{}_{j_1j_2}}{\sqrt{2!}} \\ \frac{\MA_{i_1i_2}{}^j}{\sqrt{2!}} & g_{i_1i_2,j_1j_2} + \frac{1}{2}\,\MA_{i_1i_2}{}^k\,\MA_{kj_1j_2} \end{pmatrix}.
\end{align}
We also introduce the $U$-duality-invariant tensor $\eta_{IJ;\cK}$\,, where $\cK$ denotes the index for the so-called $R_2$-representation of the $E_n$ group that can be decomposed as $(\eta_{IJ;\cK})=(\eta_{IJ;k},\,\frac{\eta_{IJ;k_1\cdots k_4}}{\sqrt{4!}})$\,. 
For $n\leq 4$, they are explicitly defined as \cite{1708.06342}
\begin{align}
 \bigl(\eta_{IJ;k}\bigr) = \begin{pmatrix} 0 & \frac{2!\,\delta^{j_1j_2}_{ik}}{\sqrt{2!}} \\ \frac{2!\,\delta^{i_1i_2}_{jk}}{\sqrt{2!}} & 0 \end{pmatrix} ,\qquad 
 \bigl(\eta_{IJ;k_1\cdots k_4}\bigr) = \begin{pmatrix} 0 & 0 \\ 0 & \frac{4!\,\delta^{i_1i_2j_1j_2}_{k_1\cdots k_4}}{\sqrt{2!\,2!}} \end{pmatrix} ,
\label{eq:eta-symbol}
\end{align}
where $\delta^{i_1\cdots i_p}_{j_1\cdots j_p}\equiv \delta^{[i_1}_{[j_1}\cdots \delta^{i_p]}_{j_p]}$\,. 
As discussed in \cite{1712.10316}, in order to formulate membrane theory in a $U$-duality-covariant manner, it is important to introduce the charge vector for a membrane
\begin{align}
 \bigl(q^\cI\bigr) \equiv \Bigl(q^i,\,\frac{q_{i_1\cdots i_4}}{\sqrt{4!}}\Bigr)\,.
\label{eq:charge-q}
\end{align}
We then define a 1-form, which we call the $\eta$-form, as
\begin{align}
 \bm{\eta}_{IJ} \equiv \eta_{IJ;\cK}\,q^\cK \,.
\end{align}
The membrane charge vector $q^\cI$ transforms covariantly under the $U$-duality transformation \eqref{eq:U-dual-J}, and accordingly, the $\eta$-form also transforms covariantly as
\begin{align}
 \bm{\eta}_{IJ} \quad \rightarrow \quad \bm{\eta}'_{IJ} = C_I{}^K\,C_J{}^L\,\bm{\eta}_{KL}\,.
\end{align}
When we consider the M2-brane (without any M5-charge induced), the charge vector should be chosen as \cite{1712.10316}
\begin{align}
 \bigl(q^\cI\bigr) = \frac{1}{2}\,\bigl(\rmd x^i,\,0\bigr) \,,
\end{align}
and then the $\eta$-form becomes
\begin{align}
 \bigl(\bm{\eta}_{IJ}\bigr) = \begin{pmatrix} 0 & \frac{\delta^{[j_1}_{i}\,\rmd x^{j_2]}}{\sqrt{2!}} \\ \frac{\delta^{[i_1}_{j}\,\rmd x^{i_2]}}{\sqrt{2!}} & 0 \end{pmatrix} \,.
\end{align}
We also define a matrix
\begin{align}
 \bigl(\bm{\cH}_I{}^J\bigr) \equiv \bigl(\bm{\eta}_{IK}\, \cH^{KJ}\bigr)
 = \begin{pmatrix} \frac{1}{2}\,\MA_{ik}{}^j\,\rmd x^k & \frac{g_{ik,j_1j_2}\,\rmd x^k+\frac{1}{2}\,\MA_{ik}{}^l\,\MA_{lj_1j_2}\,\rmd x^k}{\sqrt{2!}} \\ \frac{g^{j[i_1}\,\rmd x^{i_2]}}{\sqrt{2!}} & \frac{1}{2}\,\MA_{j_1j_2}{}^{[i_1}\,\rmd x^{i_2]} \end{pmatrix},
\end{align}
which corresponds to the matrix \eqref{eq:cHdU} defined in string theory, although here it is a 1-form. 

\paragraph{$E_n$ $U$-duality:}
By using the above definitions, we find the self-duality relation,
\begin{align}
 \cJ_I = \bm{\cH}_I{}^J\wedge * \cJ_J \,,
\label{eq:cJ-self-dual-abelian}
\end{align}
which corresponds to Eq.~\eqref{eq:self-duality-J-string}. 
Since the matrix $\bm{\cH}_I{}^J$ transforms covariantly only under the $E_n$ $U$-duality transformations, the transformation matrix $C_I{}^J$ given in Eq.~\eqref{eq:U-dual-J} should be an element of the $E_n$ group. 

Similar to the string case, we introduce 1-form fields $\cP^I(\sigma)$ as
\begin{align}
 \cJ_I \equiv \cH_{IJ}\, * \cP^J\,,\qquad 
 \bigl(\cP^I\bigr) = \begin{pmatrix} \rmd x^i \\ \frac{\MA_{i_1i_2j}\,\rmd x^j-g_{i_1i_2,j_1j_2}*(\rmd x^{j_1}\wedge\rmd x^{j_2})}{\sqrt{2!}}\end{pmatrix} . 
\label{eq:P^I-def}
\end{align}
Then, from the relation \eqref{eq:cJ-self-dual-abelian}, we obtain the self-duality relation for $\cP^I$,
\begin{align}
 \cP^I = * \bigl(\bm{\cH}^I{}_J\wedge \cP^I\bigr) \qquad \bigl[\bm{\cH}^I{}_J \equiv \cH^{IK}\,\bm{\eta}_{KJ} = \bigl(\bm{\cH}^{\rm T}\bigr)^I{}_J\bigr]\,.
\label{eq:abelian-M2-SD}
\end{align}
This corresponds to the relation \eqref{eq:self-duality-P-string} in string theory. 
Since $\cJ_I$ and $\cH_{IJ}$ transform as
\begin{align}
 \cJ'_I = C_I{}^J\,\cJ_J \,,\qquad 
 \cH'_{IJ} = C_I{}^K\,C_J{}^L\,\cH_{KL}\,,
\end{align}
under $U$-duality transformations, the 1-form $\cP^I(\sigma)$ should be transformed as
\begin{align}
 \cP'^I(\sigma) = \bigl(C^{-1}\bigr){}_J{}^I \, \cP^J(\sigma) \,.
\label{eq:P-CP}
\end{align}
The $U$-duality-covariant equations of motion $\rmd \cJ_I=0$ can be also expressed as
\begin{align}
 \rmd \bigl(\cH_{IJ}\, * \cP^J\bigr) = 0 \quad \Leftrightarrow\quad 
 \rmd * \cP^I = 0 \,.
\label{eq:eom-M-Abelian}
\end{align}

\paragraph{An issue specific to $U$-duality:}
So far, everything is parallel to the string case. 
However, as it has been pointed out in \cite{1509.02915}, the transformation \eqref{eq:P-CP} generally causes an issue. 
Here, we explain the issue by following the presentation given in \cite{1712.10316}. 
By using the self-duality relation \eqref{eq:abelian-M2-SD}, the equations of motion \eqref{eq:eom-M-Abelian} are equivalent to
\begin{align}
 \rmd \bigl(\bm{\cH}^I{}_J\wedge \cP^J\bigr) = - \bm{\cH}^I{}_J\wedge \rmd \cP^J = 0 \,.
\label{eq:eom-membrane-abelian}
\end{align}
Unlike the string case, $\bm{\cH}^I{}_J$ is not invertible and they are not equivalent to $\rmd \cP^I = 0$\,. 
To be more precisely, the equations of motion are weaker than the (Abelian) MC equation $\rmd \cP^I = 0$\,. 
Indeed, in \cite{1509.02915}, an explicit solution of membrane theory where the equations of motion \eqref{eq:eom-membrane-abelian} are satisfied but $\rmd \cP^I\neq 0$\,. 
By the definition of $\cP^I$ given in Eq.~\eqref{eq:P^I-def}, the first component $\cP^i$ trivially satisfies $\rmd \cP^i= 0$, but for the second component $\cP_{i_1i_2}$, $\rmd \cP_{i_1i_2} = 0$ is not ensured. 

Let us suppose that we have a solution $x^i(\sigma)$ satisfying $\rmd P_{i_1i_2}(\sigma)\neq 0$\,. 
Under a particular $U$-duality transformation,\footnote{The $\GL(n)$ matrix contained in the $E_n$ group has the form $\abs{\det(\Lambda_i{}^j)}^{\frac{1}{9-n}} \Bigl(\begin{smallmatrix} \Lambda_i{}^j & 0 \\ 0 & (\Lambda^{-1})_{j_1}{}^{[i_1}\,(\Lambda^{-1})_{j_2}{}^{i_2]} \end{smallmatrix}\Bigr)$, but since we are using the generalized tensors with the effective weight $0$, the determinant factor is dropped out.}
\begin{align}
 \bigl(C_I{}^J\bigr) = \begin{pmatrix} \Lambda_i{}^k & 0 \\ 0 & (\Lambda^{-1})_{k_1}{}^{[i_1}\,(\Lambda^{-1})_{k_2}{}^{i_2]} \end{pmatrix}
 \begin{pmatrix} \delta_k^j & -\frac{c_{kj_1j_2}}{\sqrt{2!}} \\ 0 & \delta_{j_1j_2}^{k_1k_2} \end{pmatrix} 
\label{eq:C-geometric}
\end{align}
with $\Lambda_i{}^j$ and $c_{ijk}$ constants, the (constant) supergravity fields are transformed as
\begin{align}
 g'_{ij} = \Lambda_i{}^k\,\Lambda_j{}^l\,g_{kl}\,,\qquad 
 \MA'_{i_1i_2i_3} = \Lambda_{i_1}{}^{j_1}\,\Lambda_{i_2}{}^{j_2}\,\Lambda_{i_3}{}^{j_3}\,\bigl(\MA_{j_1j_2j_3}+c_{j_1j_2j_3}\bigr)\,.
\end{align}
At the same time, the 1-form fields are transformed as
\begin{align}
 \bigl(\cP^I\bigr) \ \to \ \bigl(\cP'^I\bigr) = \begin{pmatrix} (\Lambda^{-1})_j{}^i\,\rmd x^j \\ 
 \frac{\MA'_{i_1i_2j}\,(\Lambda^{-1})_k{}^j\,\rmd x^k - g'_{i_1i_2,j_1j_2}\,(\Lambda^{-1})_{k_1}{}^{j_1}\,(\Lambda^{-1})_{k_2}{}^{j_2}\,* (\rmd x^{k_1}\wedge\rmd x^{k_2})}{\sqrt{2!}}\end{pmatrix}.
\end{align}
This shows that
\begin{align}
 x'^i(\sigma) = \bigl(\Lambda^{-1}\bigr)_j{}^i\, x^j(\sigma) \,,
\end{align}
is a solution of membrane theory in the dual geometry, and the (geometric) $U$-duality \eqref{eq:C-geometric} always maps a solution to the dual solution. 
On the other hand, a serious problem happens if we consider the (non-geometric) $\Omega$-transformation,
\begin{align}
 \bigl(C_I{}^J\bigr) = \begin{pmatrix} \delta_i{}^j & 0 \\ \frac{\omega^{i_1i_2j}}{\sqrt{2!}} & \delta_{j_1j_2}^{i_1i_2} \end{pmatrix}.
\label{eq:C-non-geometric}
\end{align}
After the $\Omega$-transformation, we obtain
\begin{align}
 \bigl(\cP'^I\bigr) \equiv \begin{pmatrix} \cP'^i \\ \frac{\cP'_{i_1i_2}}{\sqrt{2!}} \end{pmatrix} 
 = \begin{pmatrix} \rmd x^i + \frac{1}{2}\,\omega^{ij_1j_2}\, \cP_{j_1j_2} \\ \frac{\cP_{i_1i_2}}{\sqrt{2!}} \end{pmatrix}.
\end{align}
By assumption, we have $\rmd \cP_{i_1i_2} \neq 0$, and thus $\rmd \cP'^i(\sigma) \neq 0$\,. 
This shows that we cannot parameterize the dualized 1-form field $\cP'^I(\sigma)$ as
\begin{align}
 \bigl(\cP'^I\bigr) = \begin{pmatrix} \rmd x'^i \\ \frac{\MA'_{i_1i_2j}\,\rmd x'^j-g'_{i_1i_2,j_1j_2}*(\rmd x'^{j_1}\wedge\rmd x'^{j_2})}{\sqrt{2!}}\end{pmatrix} ,
\end{align}
because the integrability $\rmd \cP'^i=\rmd^2 x'^i(\sigma)=0$ is now violated. 

In general, $E_n$ $U$-duality transformations (for $n\leq 4$) are generated by the geometric transformations \eqref{eq:C-geometric} and the $\Omega$-transformation \eqref{eq:C-non-geometric}, but only the former preserve the integrability. 
Thus, it is concluded in \cite{1509.02915} that only the geometric subgroup of $U$-duality is the (classical) symmetry of membrane theory. 
A resolution has been discussed in \cite{1712.10316}, but even in their approach, it is impossible to realize the full MC equation $\rmd \cP^I(\sigma) = 0$.
Therefore, unlike the string case, we cannot express the 1-form as
\begin{align}
 \cP^I(\sigma) = \rmd x^I(\sigma)\,,\qquad \bigl(x^I\bigr) \equiv \begin{pmatrix} x^i \\ \frac{y_{i_1i_2}}{\sqrt{2!}} \end{pmatrix}.
\end{align}

In summary, the point is that the equations of motion $\bm{\cH}^I{}_J\wedge \rmd \cP^J = 0$ are weaker than the MC equation $\rmd \cP^I = 0$ and we cannot realize $\cP^I(\sigma) = \rmd x^I(\sigma)$ even under the equations of motion. 
Accordingly, unlike the string case, we cannot interpret that the membrane is fluctuating in an extended spacetime with coordinates $x^I$. 

\paragraph{An exceptional case where $n=3$:}
As discussed in \cite{1509.02915}, the case $n=3$ is exceptional.
There, the membrane is called the topological membrane because it is non-dynamical. 
Indeed, by using the identity
\begin{align}
 * \bigl(\rmd x^{i}\wedge\rmd x^{j}\bigr) = \varepsilon^{ij}{}_k\,\rmd x^k\qquad \Bigl(\varepsilon^{012}= \tfrac{1}{\sqrt{\abs{g}}} \Bigr)\,,
\end{align}
the equations of motion \eqref{eq:eom-Abelian-U} are identically satisfied. 
Moreover, as it is clear from
\begin{align}
 \bigl(\cP^I\bigr) = \begin{pmatrix} \rmd x^i \\ \frac{\MA_{i_1i_2j}\,\rmd x^j - \varepsilon_{i_1i_2j}\,\rmd x^j}{\sqrt{2!}}\end{pmatrix} ,
\end{align}
the (Abelian) MC equation
\begin{align}
 \rmd \cP^I(\sigma) =0\,,
\label{eq:MC-Abelian-U}
\end{align}
is also identically satisfied, and at least locally, we can express the 1-form as
\begin{align}
 \cP^I(\sigma) = \rmd x^I(\sigma) \,.
\end{align}
Here, $x^I$ describes the embedding of the membrane into the 6-dimensional extended space. 
We can freely rotate a given solution $x^I(\sigma)$ as
\begin{align}
 x^I(\sigma)\ \to\ x'^I(\sigma) = \bigl(C^{-1}\bigr){}_J{}^I \, x^J(\sigma) \,,
\end{align}
under the full $\SL(2)\times \SL(3)$ $U$-duality transformation, and here the $U$-duality group is not restricted to the geometric subgroup. 

For $n\geq 4$, only a part of $\rmd \cP^I(\sigma) =0$ can be derived from the equations of motion. 
As discussed in section \ref{sec:intro}, naively we have only $n$ equations of motion, but the number of the non-trivial components of the MC equation $\rmd \cP_{ij}=0$ are $\frac{n(n-1)}{2}$.
Therefore, these coincide only when $n=3$ (see \cite{hep-th:9512203} for a similar discussion). 
For $n\geq 4$, we cannot expect to obtain the full components of $\rmd \cP_{ij}=0$\,. 
If any component of the MC equation is not satisfied, we obtain $\rmd \cP'^i\neq 0$ after a certain $\Omega$-transformation, and the integrability is broken. 

Of course, as it is discussed well in EFT, at the level of supergravity, the Lagrangian or the equations of motion have the $E_n$ $U$-duality symmetry for an arbitrary $n\leq 8$ (or perhaps $n\leq 11$). 
The issue arises only when we try to realize the symmetry in membrane theory. 
A membrane is only a member of the supersymmetric branes, which form a $U$-duality multiplet.
In order to realize the full $U$-duality symmetry, we will need to formulate a brane theory which describes all of the supersymmetric branes in a unified manner (see section \ref{sec:discussion} for more discussion on such formulation). 
At present, such a formulation has not been found, and we can realize the $U$-duality symmetry only for the topological membrane. 
Accordingly, as we discuss below, we can realize non-Abelian $U$-duality only for the topological membrane. 

\section{Non-Abelian $T$-/$U$-duality}
\label{sec:non-Abelian}

In this section, we study the non-Abelian extension of $U$-duality. 

\subsection{PL $T$-duality in string theory}
\label{sec:PL-T}

Before studying non-Abelian $U$-duality, we review the PL $T$-duality in string theory \cite{hep-th:9502122,hep-th:9509095}. 

\paragraph{PL $T$-dualizability:}
In order to perform the PL $T$-duality, the target geometry is required to satisfy the differential equations \cite{hep-th:9502122,hep-th:9509095}
\begin{align}
 \Lie_{v_a}E_{mn}= -\tilde{f}^{bc}{}_a\,E_{mp}\,v_b^p\, v_c^q\,E_{qn} \,,
\label{eq:PL-dualizability}
\end{align}
where $E_{mn}(x)\equiv g_{mn}(x)+B_{mn}(x)$ and $v_a^m$ ($a=1,\dotsc,D$) are a set of vector fields satisfying the algebra $[v_a, v_b] = f_{ab}{}^c\, v_c$\,. 
Under this setup, the string equations of motion are expressed as the MC equation,
\begin{align}
 \rmd \cJ_a - \frac{1}{2}\,\tilde{f}_a{}^{bc}\, \cJ_b\wedge \cJ_c = 0 \,, \qquad
 \cJ_a \equiv v_a^m \,\bigl(g_{mn} *\rmd x^n + B_{mn}\, \rmd x^n\bigr) \,.
\label{eq:J_a-string}
\end{align}

As discussed in \cite{hep-th:9502122,hep-th:9509095}, Eq.~\eqref{eq:PL-dualizability} suggests that $f_{ab}{}^c$ and $\tilde{f}^{ab}{}_c$ can be identified with the structure constants of the Lie algebra of the Drinfel'd double,
\begin{align}
 [T_a,\,T_b] = f_{ab}{}^c\,T_c\,,\qquad
 [\tilde{T}^a,\,\tilde{T}^b] = \tilde{f}^{ab}{}_c\,\tilde{T}^c\,, \qquad
 [T_a,\,\tilde{T}^b] = \tilde{f}^{bc}{}_a\,T_c - f_{ac}{}^b\,\tilde{T}^c\,.
\label{eq:Drinfeld}
\end{align}
This is sometimes expressed as $[T_A,\,T_B]=\cF_{AB}{}^C\,T_C$ by denoting the set of generators as $\{T_A\}\equiv \{T_a,\,\tilde{T}^a\}$\,. 
In addition, an ad-invariant\footnote{The ad-invariance means $\langle [T_C,\,T_A],\,T_B\rangle + \langle T_A,\,[T_C,\,T_B]\rangle=0$\,.} bilinear form is defined for the generators,
\begin{align}
 \langle T_A,\,T_B\rangle \equiv \eta_{AB}\,,\qquad 
 \bigl(\eta_{AB}\bigr)\equiv \begin{pmatrix} 0 & \delta_a^b \\ \delta^a_b & 0 \end{pmatrix} .
\label{eq:Drinfeld-bilinear}
\end{align}

We denote a subgroup $G$ generated by $\{T_a\}$ as the physical subgroup while a subgroup $\tilde{G}$ generated by $\{\tilde{T}^a\}$ as the dual group. 
If we assume that the target space is a group manifold of $G$ and identify the vector fields $v_a^m$ with the left-invariant vector fields, we can solve the differential equation \eqref{eq:PL-dualizability} as follows \cite{hep-th:9502122}:
\begin{align}
 E_{mn}(x) = \bigl[\hat{E}\,(\bm{1}-\Pi(x)\,\hat{E})^{-1}\bigr]_{ab}\, r^a_m(x)\, r^b_n(x) \,.
\label{eq:PL-solution}
\end{align}
Here, $\hat{E}\equiv (\hat{E}_{ab})$ is a constant matrix and several quantities are defined as follows. 
For a group element $g(x)\in G$, we define a matrix $M_A{}^B(x)$ as
\begin{align}
 g^{-1}\, T_A\, g \equiv M_A{}^B\, T_B \,. 
\label{eq:M-def1}
\end{align}
From the structure of the algebra \eqref{eq:Drinfeld}, the matrix $M_A{}^B(x)$ can be generally parameterized as follows by using two matrices $a_a{}^b(x)$ and $\Pi^{ab}(x)=-\Pi^{ba}(x)$:
\begin{align}
 \bigl(M_A{}^B\bigr) \equiv \begin{pmatrix} \delta_a^c & 0 \\ -\Pi^{ac} & \delta^a_c \end{pmatrix} \begin{pmatrix} a_c{}^b & 0 \\ 0 & (a^{-1})_b{}^c \end{pmatrix} .
\label{eq:M-def2}
\end{align}
The left- and right-invariant 1-forms are denoted as
\begin{align}
 \ell \equiv g^{-1}\,\rmd g\,,\qquad r \equiv \rmd g\,g^{-1} \qquad \bigl(\ell^a_m = a_b{}^a \,r^{b}_m\bigr) \,,
\end{align}
and they are dual to the left- and right-invariant vectors $v_a^m$ and $e_a^m$,
\begin{align}
 \ell^a_m\,v^m_b = \delta^a_b\,,\qquad r^a_m\,e^m_b = \delta^a_b\,.
\end{align}
Now, the solution \eqref{eq:PL-solution} can be understood from these definitions. 
When a target geometry takes the form of Eq.~\eqref{eq:PL-solution}, we can perform the PL $T$-duality. 

\paragraph{Manifest $T$-duality:}
The solution \eqref{eq:PL-solution} can be rewritten in a nicer form by using the generalized metric $\cH_{MN}$. 
Indeed, the solution \eqref{eq:PL-solution} can be expressed as
\begin{align}
 \cH_{MN}(x) = E_M{}^A(x)\,E_N{}^B(x)\, \hat{\cH}_{AB} \,.
\label{eq:cH-double}
\end{align}
Here, $\hat{\cH}_{AB}$ is a constant matrix associated with $\hat{E}_{ab}\equiv (\hat{g}+\hat{B})_{ab}$ ($\hat{g}_{ab}\equiv \hat{E}_{(ab)}$, $\hat{B}_{ab}\equiv \hat{E}_{[ab]}$) and the coordinate dependence is contained only in the twist matrix $E_M{}^A(x)$,
\begin{align}
 \bigl(\hat{\cH}_{AB}\bigr) \equiv \begin{pmatrix} (\hat{g} - \hat{B}\,\hat{g}^{-1}\,\hat{B})_{ab} & (\hat{B}\,\hat{g}^{-1})_a{}^b \\ -(\hat{g}^{-1}\,\hat{B})_a{}^b & \hat{g}^{ab} \end{pmatrix},\qquad
 \bigl(E_M{}^A\bigr) \equiv \begin{pmatrix} r_m^a & 0 \\ -e^m_c\,\Pi^{c a} & e^m_a \end{pmatrix}. 
\end{align}
We note that the inverse of the twist matrix, denoted as $E_A{}^M$, is known as the generalized frame fields and, in fact, they satisfy the relation \cite{1707.08624}
\begin{align}
 \gLie_{E_A}E_B{}^M = - \cF_{AB}{}^C\,E_C{}^M\,,
\label{eq:gen-frame-Lie}
\end{align}
where $\gLie$ denotes the generalized Lie derivative in DFT. 
Then, we can show that the generalized metric satisfies the equation
\begin{align}
 \gLie_{E_A} \cH_{MN} = \cF_{AM}{}^P\,\cH_{PN} + \cF_{AN}{}^P\,\cH_{MP}\qquad \bigl(\cF_{AM}{}^N\equiv \cF_{AB}{}^C\,E_M{}^B\,E_C{}^N\bigr)\,,
\end{align}
which shows that the target space has the symmetry of the Drinfel'd double. 

Now, we rewrite the equations of motion \eqref{eq:J_a-string} into a $T$-duality-manifest form. 
Similar to the Abelian case, we define 1-form fields
\begin{align}
 \bigl(P^A\bigr) = \begin{pmatrix} \ell^a \\ \cJ_a \end{pmatrix}
 = \begin{pmatrix} \ell^a_m\,\rmd x^m \\ v_a^m \,\bigl(g_{mn} *\rmd x^n + B_{mn}\, \rmd x^n\bigr) \end{pmatrix},
\end{align}
which reduce to Eq.~\eqref{eq:string-P} in the Abelian case (where $\ell^a_m=\delta^a_m$ and $v_a^m=\delta_a^m$). 
For convenience, we also define
\begin{align}
 P(\sigma)\equiv P^A(\sigma)\,T_A = \ell + \cJ \qquad \bigl(\ell \equiv \ell^a\,T_a\,,\quad \cJ\equiv \cJ_a\,\tilde{T}^a\bigr) \,. 
\end{align}
By further acting the adjoint action, we define
\begin{align}
\begin{split}
 &\cP(\sigma)\equiv \cP^A\,T_A \equiv g\,P(\sigma)\,g^{-1} = g\, (\ell + \cJ)\, g^{-1}
\\
 &\Biggl[\Leftrightarrow\ \cP^A=(M^{-1})_B{}^A\,P^B = E_M{}^A\,P^M\,,\qquad 
 \bigl(P^M\bigr) \equiv \begin{pmatrix} \rmd x^m \\ g_{mn} *\rmd x^n + B_{mn}\, \rmd x^n \end{pmatrix} \Biggr]\,.
\end{split}\label{eq:cP-def-T}
\end{align}
Eq.~\eqref{eq:J_a-string} suggests that, under the equations of motion, $\cJ$ can be identified with the right-invariant 1-form $\tilde{r} \equiv \rmd \tilde{g}\,\tilde{g}^{-1}$ associated with a dual group element $\tilde{g}(\tilde{x})$, and we obtain
\begin{align}
 \cP(\sigma) = g\, \bigl(g^{-1} \rmd g + \rmd \tilde{g}\,\tilde{g}^{-1}\bigr)\, g^{-1} = \rmd l\,l^{-1}\qquad \bigl(l\equiv g\,\tilde{g}\bigr)\,.
\end{align}
This shows that the 1-form field $\cP(\sigma)$ is the right-invariant 1-form on the Drinfel'd double, which satisfies
\begin{align}
 \rmd \cP - \cP\wedge\cP =0 \qquad\text{ or }\qquad\rmd \cP^A - \frac{1}{2}\,\cF_{BC}{}^A\,\cP^B\wedge\cP^C = 0 \,.
\label{eq:Drinfeld-MC}
\end{align}
Similar to the Abelian case, the 1-form fields are subjected to the self-duality relation,
\begin{align}
 \cP^A = \cH^A{}_B \,*\cP^B \qquad \bigl(\cH^A{}_B\equiv \hat{\cH}^{AC}\,\eta_{CB}\bigr)\,,
\label{eq:string-SD}
\end{align}
and only $D$ components are independent. 
Thus, the $2D$ MC equation \eqref{eq:Drinfeld-MC} are equivalent to the $D$ equations of motion given in Eq.~\eqref{eq:J_a-string}. 

\paragraph{PL $T$-duality:}
The PL $T$-duality (or the PL $T$-plurality \cite{hep-th:0205245}) is a symmetry under redefinitions of the generators
\begin{align}
 T'_A = C_A{}^B\,T_B \,.
\label{eq:PL-ODD}
\end{align}
Under the redefinition, the structure constants are transformed as
\begin{align}
 \cF'_{AB}{}^C = C_A{}^D\,C_B{}^E\, \bigl(C^{-1}\bigr){}_F{}^C\,\cF_{DE}{}^F\,. 
\end{align}
By requiring that the redefined algebra is also a Lie algebra of the Drinfel'd double, the metric $\eta_{AB}$ [i.e.~the bilinear form \eqref{eq:Drinfeld-bilinear}] must be preserved
\begin{align}
 C_A{}^C\,C_B{}^D\,\eta_{CD}=\eta_{AB} \,.
\end{align}
Namely, the constant matrix $C_A{}^B$ should be an element of the $\OO(D,D)$ group. 
After the redefinition, we introduce new group elements $g'(\sigma)$ and $\tilde{g}'(\sigma)$, such that $g\,\tilde{g} = l=l' = g'\,\tilde{g}'$ is satisfied. 
Then, we obtain $\cP(\sigma) = \rmd l\,l^{-1} = \rmd l'\,l'^{-1} \equiv \cP'(\sigma) $, or equivalently,
\begin{align}
 \cP'^A = \bigl(C^{-1}\bigr){}_B{}^A\,\cP^B \,.
\label{eq:cP-dual}
\end{align}
This shows that the equations of motion \eqref{eq:Drinfeld-MC} and \eqref{eq:string-SD} are covariantly transformed if $\cH_{AB}$ is also transformed as
\begin{align}
 \cH'_{AB} = C_A{}^C\,C_B{}^D\, \cH_{CD} \,.
\label{eq:HAB-prime}
\end{align}
In this sense, the PL $T$-duality is an $\OO(D,D)$ transformation that covariantly transforms the equations of motion of string theory. 

In summary, the essential point of the PL $T$-duality is that the string equations of motion are expressed as the MC equation \eqref{eq:Drinfeld-MC} for a 1-form $\cP(\sigma)$ that satisfy the self-duality relation \eqref{eq:string-SD}. 
These equations are manifestly covariant under $\OO(D,D)$ PL $T$-duality transformations, given in Eqs.~\eqref{eq:PL-ODD}, \eqref{eq:cP-dual}, and \eqref{eq:HAB-prime}. 

\paragraph{Dual solution:}
Although the equations of motion are manifestly covariant, the procedure to obtain the dual string solution may be rather complicated. 
For the explicit computation, we need to fix the parameterizations of the group elements (e.g.~$g(x)=\Exp{x^aT_a}$ and $\tilde{g}(\tilde{x})=\Exp{\tilde{x}_aT^a}$). 
Given these, we can compute the original target geometry by using Eq.~\eqref{eq:PL-solution}. 
After an $\OO(D,D)$ rotation, we again provide parameterizations of group elements, such as $g'(x')=\Exp{x'^aT'_a}$ and $\tilde{g}'(\tilde{x}')=\Exp{\tilde{x}'_aT'^a}$, and then obtain the dual geometry
\begin{align}
 E'_{mn}(x) = \bigl[\hat{E}'\,(\bm{1}-\Pi'\,\hat{E}')^{-1}\bigr]_{ab}\, r'^a_m\, r'^b_n \,.
\end{align}
In order to relate the two geometries, we require 
\begin{align}
 g(x)\,\tilde{g}(x) = l = g'(x')\,\tilde{g}'(x')\,.
\end{align}
Then, in principle, we can find the relation between the two coordinates,
\begin{align}
 x'^a = x'^a\bigl(x^a,\,\tilde{x}_a\bigr)\,,\qquad 
 \tilde{x}'_a = \tilde{x}'_a\bigl(x^a,\,\tilde{x}_a\bigr)\,.
\label{eq:x-relation-string}
\end{align}
Using this relation, we can map a string solution in the original geometry to the dual solution. 
From a given solution $x^a(\sigma)$, we can compute the 1-form $\cJ_a$ defined in Eq.~\eqref{eq:J_a-string}. 
Then, solving the differential equations $\cJ = \rmd \tilde{g}\,\tilde{g}^{-1}$, we find $\tilde{x}_a(\sigma)$\,. 
Finally, substituting the solutions $x^a(\sigma)$ and $\tilde{x}_a(\sigma)$ into Eq.~\eqref{eq:x-relation-string}, we obtain the dual solution $x'^a(\sigma)$\,. 

Another easier method is as follows. 
From a given solution $x^m(\sigma)$, we can easily compute the 1-form field $\cP(\sigma)$. 
Expanding $\cP$ by means of the redefined generators $T'_A$, we obtain $\cP'^A$\,. 
Then, we can compute $P'^A =M'_A{}^B\,\cP'^A$\,, whose first component is $P'^a=\ell'^a$\,. 
Solving the differential equation $P'^a\,T'_a=g'^{-1}(x')\,\rmd g'(x')$, we can find $x'^m(\sigma)$\,. 

Either way, we can map a solution $x^m(\sigma)$ to a new solution $x'^m(\sigma)$ of the dual sigma model. 

\subsection{Non-Abelian $U$-duality}
\label{sec:NAUD}

Here, we study the $U$-duality extension of the PL $T$-duality in membrane theory. 
We note that our analysis is restricted to $n\leq 4$\,. 

\paragraph{The $\cE_n$ algebra:}
The PL $T$-duality is based on the Lie algebra of the Drinfel'd double, and similarly, non-Abelian $U$-duality will be based on a new algebra that extends the Lie algebra of the Drinfel'd double. 
Such an algebra has been recently proposed in \cite{1911.06320,1911.07833} and we call it the $\cE_n$ algebra by following \cite{1911.06320}. 
For $n\leq 4$, the algebra is given by
\begin{align}
\begin{split}
 T_a\circ T_b &= f_{ab}{}^c\,T_c \,,
\\
 T_a\circ T^{b_1b_2} &= f_a{}^{b_1b_2c}\,T_c + 2\,f_{ac}{}^{[b_1}\, T^{b_2]c} \,,
\\
 T^{a_1a_2}\circ T_b &= -f_b{}^{a_1a_2 c}\,T_c + 3\,f_{[c_1c_2}{}^{[a_1}\,\delta^{a_2]}_{b]} \,T^{c_1c_2}\,,
\\
 T^{a_1a_2}\circ T^{b_1b_2} &= -2\, f_d{}^{a_1a_2[b_1}\, T^{b_2]d}\,,
\end{split}
\label{eq:En-algebra}
\end{align}
where $a,b=1,\dotsc,n$\,. 
The indices of the generators are antisymmetric $T^{ab}=-T^{ba}$ and the structure constants have symmetries $f_{ab}{}^c=f_{[ab]}{}^c$ and $f_a{}^{b_1b_2b_3}=f_a{}^{[b_1b_2b_3]}$\,. 
For simplicity, we denote the algebra as
\begin{align}
 T_A \circ T_B = \cF_{AB}{}^C\,T_C\,,\qquad \bigl(T_A\bigr)\equiv \bigl(T_a,\,\tfrac{T^{a_1a_2}}{\sqrt{2!}}\bigr)\,.
\end{align}
Since the first two indices of the structure constants are not antisymmetric (i.e.~$\cF_{AB}{}^C\neq -\cF_{BA}{}^C$), this is a Leibniz algebra rather than a Lie algebra. 
The Leibniz identity,
\begin{align}
 T_A \circ \bigl(T_B \circ T_C\bigr) = \bigl(T_A \circ T_B\bigr) \circ T_C + T_B \circ \bigl(T_A \circ T_C\bigr)\,,
\end{align}
requires that the structure constants should satisfy \cite{1911.06320}
\begin{align}
 0&=f_{[ab}{}^{e}\,f_{c]e}{}^d\,,
\\
 0&=f_{bc}{}^e\,f_{e}{}^{a_1a_2 d} + 6\, f_{e[b}{}^{[d}\, f_{c]}{}^{a_1a_2]e} \,,
\label{eq:Leibniz-2}
\\
 0&=f_{d_1d_2}{}^{[a_1}\, \delta^{a_2]}_b\, f_c{}^{d_1d_2 e} \,,
\label{eq:Leibniz-3}
\\
 0&= 3\,f_{[d_1d_2}{}^{[a_1}\,\delta^{a_2]}_{e]}\,f_c{}^{e b_1b_2}
 + 4\,f_{ef}{}^{[a_1}\,f_c{}^{a_2]e[b_1}\,\delta^{b_2]f}_{d_1d_2} \,,
\label{eq:Leibniz-4}
\\
 0&=f_{c}{}^{e a_1a_2}\,f_e{}^{db_1b_2} - 3\, f_c{}^{e[b_1b_2}\, f_e{}^{d] a_1a_2} \,.
\end{align}
Similar to the Drinfel'd double, a $U$-duality-invariant metric has been defined as
\begin{align}
 \langle T_A,\, T_B \rangle_\cC = \eta_{AB;\cC}\,,
\end{align}
where $(\eta_{AB;\cC})\equiv \bigl(\eta_{AB;c},\,\tfrac{\eta_{AB;c_1\cdots c_4}}{\sqrt{4!}}\bigr)$ has the same matrix elements as Eq.~\eqref{eq:eta-symbol}. 

\paragraph{Target space with the $\cE_n$ symmetry:}
Similar to the case of the PL $T$-duality, where the target geometry is expressed as \eqref{eq:cH-double}, we can construct a target geometry as \cite{1911.06320}
\begin{align}
 \cH_{IJ}(x) = E_I{}^A(x)\,E_J{}^B(x)\, \hat{\cH}_{AB} \,, 
\label{eq:En-geometry1}
\end{align}
by using the $\cE_n$ algebra. 
Here, $\hat{\cH}_{AB}$ is a constant matrix and $E_I{}^A$ has the form
\begin{align}
 \bigl(E_I{}^A\bigr) \equiv \begin{pmatrix} r_i^a & 0 \\ \frac{e^{i_1}_{b_1}\,e^{i_2}_{b_2}\,\Pi^{b_1b_2a}}{\sqrt{2!}} & e^{[i_1}_{[a_1}\,e^{i_2]}_{a_2]} \end{pmatrix}. 
\end{align}
The right-invariant 1-form $r^a_i$ and its dual $e_a^i$ has been defined by using a physical group element $g(x)$, which we parameterize as $g(x)=\Exp{x^a T_a}$.
In addition, similar to Eqs.~\eqref{eq:M-def1} and \eqref{eq:M-def2}, the tri-vector $\Pi^{abc}$ is defined as
\begin{align}
\begin{split}
 &g^{-1}(x) \circ T_A \equiv M_A{}^B(x)\, T_B \,,
\\
 &\bigl(M_A{}^B\bigr) \equiv \begin{pmatrix} a_a{}^b & 0 \\ -\frac{\Pi^{a_1a_2 c}\,a_c{}^b}{\sqrt{2!}} & (a^{-1})_{[b_1}{}^{a_1}\,(a^{-1})_{b_2]}{}^{a_2} 
\end{pmatrix},
\end{split}
\end{align}
where $g^{-1}\circ T_A = \Exp{-h}\circ\, T_A$ ($h\equiv x^a\,T_a$) denotes
\begin{align}
 g^{-1}\circ T_A \equiv T_A - h\circ T_A + \frac{1}{2!}\, h\circ \bigl(h\circ T_A\bigr) - \frac{1}{3!}\, h\circ \bigl(h\circ \bigl(h\circ T_A\bigr)\bigr) + \cdots\,.
\end{align}
Similar to the case of the Drinfel'd double, the generalized frame fields $E_A{}^I$ (defined as the inverse of $E_I{}^A$) satisfy the relation \eqref{eq:gen-frame-Lie} by means of the generalized Lie derivative in EFT, and we can show that the target geometry has the generalized isometry group, which is generated by the $\cE_n$ algebra
\begin{align}
 \gLie_{E_A} \cH_{IJ} = \cF_{AI}{}^K\,\cH_{KJ} + \cF_{AJ}{}^K\,\cH_{IK}\qquad \bigl(\cF_{AI}{}^J\equiv \cF_{AB}{}^C\,E_I{}^B\,E_C{}^J\bigr)\,.
\end{align}

If we introduce the dual metric $\tilde{g}_{ij}$ and $\Omega^{ijk}$ through the non-geometric parameterization of the generalized metric (see for example \cite{1205.6403,1612.08738,1701.07819})
\begin{align}
 \bigl(\cH_{IJ}\bigr) = \begin{pmatrix} \delta_i^k & 0 \\ \frac{\Omega^{i_1i_2 k}}{\sqrt{2!}} & \delta^{i_1i_2}_{k_1k_2} \end{pmatrix} 
 \begin{pmatrix} \tilde{g}_{kl} & 0 \\ 0 & \tilde{g}^{k_1k_2,l_1l_2} \end{pmatrix} 
 \begin{pmatrix} \delta^l_j & \frac{\Omega^{lj_1j_2}}{\sqrt{2!}} \\ 0 & \delta_{l_1l_2}^{j_1j_2} \end{pmatrix} ,
\end{align}
we find
\begin{align}
 \tilde{g}_{ij}\equiv e_i^a\,e_j^b\,\hat{g}_{ab} = v_i^a\,v_j^b\,\hat{g}_{ab}\,,\qquad \Omega^{ijk}\equiv \Omega^{abc}\,e_a^i\,e_b^j\,e_c^k \quad \bigl(\Omega^{abc}\equiv \Pi^{abc}+\hat{\Omega}^{abc}\bigr) \,,
\label{eq:dual-fields}
\end{align}
where we have parameterized the constant matrix $\hat{\cH}_{AB}$ as
\begin{align}
 \bigl(\hat{\cH}_{AB}\bigr) = \begin{pmatrix} \delta_a^c & 0 \\ \frac{\hat{\Omega}^{a_1a_2 c}}{\sqrt{2!}} & \delta^{a_1a_2}_{c_1c_2} \end{pmatrix} 
 \begin{pmatrix} \hat{g}_{cd} & 0 \\ 0 & \hat{g}^{c_1c_2,d_1d_2} \end{pmatrix} 
 \begin{pmatrix} \delta^d_b & \frac{\hat{\Omega}^{db_1b_2}}{\sqrt{2!}} \\ 0 & \delta_{d_1d_2}^{b_1b_2} \end{pmatrix}\,,
\end{align}
where $\hat{\Omega}^{abc}$ and $\hat{g}_{ab}$ are constants that are assumed to satisfy
\begin{align}
 f_{a(b}{}^d\,\hat{g}_{c)d}=0\,,\qquad f_{de}{}^{[a}\,\hat{\Omega}^{bc]e} =0\,.
\label{eq:bi-invariance}
\end{align}
Then, we find that the dual fields satisfy \cite{1911.06320}
\begin{align}
 \Lie_{v_a} \tilde{g}_{ij}=0\,,\qquad \Lie_{v_a} \Omega^{ijk}=f_a{}^{bcd}\,v_b^i\,v_c^j\,v_d^k\,.
\label{eq:Lie-dual}
\end{align}
The target space constructed in this way is the setup to discuss non-Abelian $U$-duality. 

In order to identify the standard supergravity fields, we make the following identification between the standard fields $(g_{ij},\,\MA_{ijk})$ and the dual fields $(\tilde{g}_{ij},\,\Omega^{ijk})$ \cite{1205.6403,1612.08738} (see also \cite{Duff:1990hn}):
\begin{align}
\begin{split}
 &\abs{\tilde{g}}^{\frac{1}{9-n}}
\begin{pmatrix} \delta_i^k & 0 \\ \frac{\Omega^{i_1i_2 k}}{\sqrt{2!}} & \delta^{i_1i_2}_{k_1k_2} \end{pmatrix} 
 \begin{pmatrix} \tilde{g}_{kl} & 0 \\ 0 & \tilde{g}^{k_1k_2,l_1l_2} \end{pmatrix} 
 \begin{pmatrix} \delta^l_j & \frac{\Omega^{lj_1j_2}}{\sqrt{2!}} \\ 0 & \delta_{l_1l_2}^{j_1j_2} \end{pmatrix}
\\
 &=
 \abs{g}^{\frac{1}{9-n}} \begin{pmatrix} \delta_i^k & -\frac{\MA_{ik_1k_2}}{\sqrt{2!}} \\ 0 & \delta^{i_1i_2}_{k_1k_2} \end{pmatrix} 
 \begin{pmatrix} g_{kl} & 0 \\ 0 & g^{k_1k_2,l_1l_2} \end{pmatrix} 
 \begin{pmatrix} \delta^l_j & 0 \\ -\frac{\MA_{l_1l_2j}}{\sqrt{2!}} & \delta_{l_1l_2}^{j_1j_2} \end{pmatrix},
\end{split}
\end{align}
where the density factors are needed in order to remove the weight of the generalized metric. 
From this relation, the standard supergravity fields (for $n\leq 4$) are obtained as follows:
\begin{align}
 g_{ij} = K^{\frac{2}{3}}\,\Bigl(K^{-1}\,\tilde{g}_{ij} - \frac{1}{2}\,\Omega_{ikl}\,\Omega^{kl}{}_j\Bigr)\,, \qquad 
 \MA_{ijk} = -K\,\Omega_{ijk} \,,
\label{eq:En-geometry2}
\end{align}
where $K^{-1} \equiv 1+\frac{1}{3!}\,\Omega^{ijk}\,\Omega_{ijk}$ and the indices of $\MA_{ijk}$ and $\Omega^{ijk}$ are raised or lowered by the metric $g_{ij}$ and the dual metric $\tilde{g}_{ij}$, respectively. 

In terms of the standard fields, the relations \eqref{eq:Lie-dual} read\footnote{They can be checked by using relations specific to $n=3,4$ given in Appendices \ref{app:n=4} and \ref{app:n=3}.}
\begin{align}
\begin{split}
 \Lie_{v_a}g_{ij} &= - \frac{2}{3\cdot 3!}\,f_a{}^{bcd}\,\MA_{bcd}\,g_{ij} + f_a{}^{bcd}\,\MA_{bc(i}\,g_{j)d}\,,
\\
 \Lie_{v_a}\MA_3 &= - \frac{1}{3!}\,f_a{}^{bcd}\,\ell_b\wedge\ell_c\wedge\ell_d + \frac{1}{3!}\,f_a{}^{bcd}\,\MA_{bcd}\,\MA_3\,,
\end{split}
\label{eq:Lie-geometric}
\end{align}
where $\ell_a\equiv g_{ab}\,\ell^b$, and curved indices $i,j$ of $g_{ij}$ and $\MA_{ijk}$ have been converted to the indices $a,b$ by using $v_a^i$ (e.g.~$\MA_{abc}\equiv v_a^i\,v_b^j\,v_c^k\,\MA_{ijk}$). 

We also note that the metric $G_{ab}\equiv e_a^i\,e_b^j\,g_{ij}$ also satisfies
\begin{align}
 f_{a(b}{}^d\,G_{c)d}=0\,.
\label{eq:invariant-metric}
\end{align}
Indeed, in $n=3$, $g_{ij}\propto \tilde{g}_{ij}$ and Eq.~\eqref{eq:invariant-metric} is trivial. 
In $n=4$, we can parameterize $\Omega^{ijk}$ as $\Omega^{ijk}=\tilde{\varepsilon}^{ijkl}\,\Omega_l$ $\bigl(\tilde{\varepsilon}^{0123}=\frac{1}{\sqrt{\abs{\tilde{g}}}}\bigr)$ (see Appendix \ref{app:n=4}) and then we obtain
\begin{align}
 g_{ij} = K^{\frac{2}{3}}\,\bigl(\tilde{g}_{ij}-\Omega_i\,\Omega_j\bigr)\,,\qquad 
 K=\frac{1}{1-\tilde{g}^{ij}\,\Omega_i\,\Omega_j}\,. 
\end{align}
By assuming $f_{ab}{}^a=0$, Eq.~\eqref{eq:bi-invariance} leads to $f_{ab}{}^c\,\hat{\Omega}_c=0$ where $\hat{\Omega}^{abc}\equiv \tilde{\varepsilon}^{abcd}\,\hat{\Omega}_d$\,. 
Moreover, from the identity \eqref{eq:id3}, we obtain $f_{ab}{}^c\,(\Omega^{abd}-\hat{\Omega}^{abd}) =0$\,, which is equivalent to $f_{[ab}{}^d\, (\Omega_{c]}-\hat{\Omega}_{c]}) =0$.
Then, we obtain
\begin{align}
 f_{ab}{}^c\, \Omega_{c} = f_{ab}{}^c\, \bigl(\Omega_{c}-\hat{\Omega}_c\bigr) = - 2\,f_{c[a}{}^c\, \bigl(\Omega_{b]}-\hat{\Omega}_{b]}\bigr)=0\,.
\end{align}
This shows the desired relation in $n=4$,
\begin{align}
 f_{a(b}{}^d\,G_{c)d} = K^{\frac{2}{3}}\,\bigl(f_{a(b}{}^d\,\tilde{g}_{c)d}-f_{a(b}{}^d\,\Omega_{c)}\,\Omega_d\bigr) =0 \,.
\end{align}
Thus, both in $n=3$ and $n=4$ (with $f_{ab}{}^a=0$), $G_{ab}$ is an invariant metric, and we have
\begin{align}
 G_{ab} = \bigl(a^{-1}\bigr){}_a{}^c\,\bigl(a^{-1}\bigr){}_b{}^d\,G_{cd} = v_a^i\,v_b^j\,g_{ij} = g_{ab} \,.
\end{align}

\paragraph{Membrane theory:}
Now, let us consider membrane theory. 
In a general curved spacetime, the equations of motion for the scalar fields $x^i(\sigma)$ become
\begin{align}
 &\partial_{\WSa}\Bigl(\sqrt{-h}\,g_{ij}\,h^{\WSa\WSb}\,\partial_{\WSb}x^j + \frac{1}{2}\,\MA_{ijk}\,\epsilon^{\WSa\WSb\WSc}\,\partial_{\WSb}x^j\,\partial_{\WSc}x^k \Bigr)
\nn\\
 &= \frac{1}{2}\,\partial_i g_{jk}\,h^{\WSa\WSb}\,\partial_{\WSa}x^k\,\partial_{\WSb}x^l 
 + \frac{1}{3!}\,\partial_i \MA_{k_1k_2k_3}\,\epsilon^{\WSa\WSb\WSc}\,\partial_{\WSa}x^{k_1}\,\partial_{\WSb}x^{k_2}\,\partial_{\WSc}x^{k_3} \,,
\end{align}
where $\epsilon^{012}=1$\,. 
By contracting the free index $i$ with a set of vector fields $v_a^i$, we obtain
\begin{align}
 \rmd \cJ_a = \frac{1}{2}\,\Lie_{v_a}g_{ij}\,\rmd x^i\wedge *\rmd x^j - \Lie_{v_a}\MA_3 \,, 
\end{align}
where we have defined
\begin{align}
 \cJ_a \equiv * \ell_a - \frac{1}{2}\,\MA_{abc}\,\ell^b\wedge\ell^c \,. 
\end{align}
In the target geometry given by Eqs.~\eqref{eq:En-geometry1} and \eqref{eq:En-geometry2}, by choosing the vector fields $v_a^i$ as the left-invariant vector fields, the equations of motion become
\begin{align}
 \rmd \cJ_a = \frac{1}{3!}\,f_a{}^{bcd}\, \bigl[ \ell_b\wedge\ell_c\wedge\ell_d + \bigl(3\,\MA_{bce}\,H^e_d - \MA_{bcd}\bigr) *1 - \MA_{bcd}\,\MA_3\bigr] \,,
\label{eq:eom-En}
\end{align}
where $H^a_b\equiv h^{\WSa\WSb}\,\ell^a_i\,\ell^c_j\,g_{cb}\,\partial_{\WSa}x^i\,\partial_{\WSb}x^j$\,. 
Note that $H^a_b$ is a projector satisfying $H^a_c\,H^c_b = H^a_b$ in $n=4$ while $H^a_b=\delta^a_b$ in $n=3$. 
Note also that Eq.~\eqref{eq:eom-En} reduces to the equations of motion \eqref{eq:eom-Abelian-U} in the Abelian case, where $f_a{}^{bcd}=0$ and $\ell^a = \delta^a_i\,\rmd x^i$\,. 

For the manifest $U$-duality, we define a combination 
\begin{align}
 \bigl(\cJ_A\bigr) \equiv \begin{pmatrix} \cJ_a \\ \frac{\ell^{a_1}\wedge\ell^{a_2}}{\sqrt{2!}} \end{pmatrix} ,
\end{align}
similar to the Abelian case \eqref{eq:cJ-Abelian-U}. 
Similar to Eq.~\eqref{eq:P^I-def}, we also define the Hodge dual of the 2-form field $\cJ_A$ through
\begin{align}
 \cJ_A = \cH_{AB}\, * P^B\,, 
\end{align}
where
\begin{align}
 \bigl(\cH_{AB}\bigr) \equiv \begin{pmatrix} \delta_a^c & -\frac{\MA_{ac_1c_2}}{\sqrt{2!}} \\ 0 & \delta^{a_1a_2}_{c_1c_2} \end{pmatrix} 
 \begin{pmatrix} g_{cd} & 0 \\ 0 & g^{c_1c_2,d_1d_2} \end{pmatrix} 
 \begin{pmatrix} \delta^d_b & 0 \\ -\frac{\MA_{d_1d_2b}}{\sqrt{2!}} & \delta_{d_1d_2}^{b_1b_2} \end{pmatrix}.
\end{align}
Then, we find that the 1-form fields have the form
\begin{align}
 \bigl(P^A\bigr) \equiv \begin{pmatrix} P^a \\ \frac{P_{a_1a_2}}{\sqrt{2!}}
 \end{pmatrix} = \begin{pmatrix} \ell^a \\ \frac{\MA_{a_1a_2b}\,\ell^b - *(\ell_{a_1}\wedge\ell_{a_2})}{\sqrt{2!}}
 \end{pmatrix}. 
\end{align}
Similar to the case of the PL $T$-duality [see Eq.~\eqref{eq:cP-def-T}], we redefine the 1-form fields as
\begin{align}
 \cP \equiv \cP^A\,T_A = g\circ \Bigl(P^a\,T_a + \frac{1}{2}\,P_{a_1a_2}\,T^{a_1a_2}\Bigr)\quad \Leftrightarrow\quad \cP^A=\bigl(M^{-1}\bigr)_B{}^A\, P^B\,.
\label{eq:cP-def}
\end{align}
Then, we obtain
\begin{align}
 \bigl(\cP^A\bigr) \equiv \begin{pmatrix} \cP^a \\ \frac{\cP_{a_1a_2}}{\sqrt{2!}}
 \end{pmatrix} = \begin{pmatrix} r^a + \frac{1}{2}\,\Pi^{ab_1b_2}\,\cP_{b_1b_2} \\ \frac{a_{a_1}{}^{b_1}\,a_{a_2}{}^{b_2}\,P_{b_1b_2}}{\sqrt{2!}}
 \end{pmatrix} .
\label{eq:cP-parameterization}
\end{align}
Similar to Eq.~\eqref{eq:string-SD}, this satisfies the self-duality relation
\begin{align}
 \cP^A = * \bigl(\bm{\hat{\cH}}{}^A{}_B\wedge \cP^B\bigr) \qquad \bigl(\bm{\hat{\cH}}{}^A{}_B \equiv \hat{\cH}^{AC}\,\bm{\eta}_{CB} \bigr)\,,
\label{eq:SD-NAUD}
\end{align}
where we have defined the $\eta$-form as
\begin{align}
 \bigl(\bm{\eta}_{AB}\bigr) \equiv \begin{pmatrix} 0 & \frac{\delta^{[b_1}_{a}\,\ell^{b_2]}}{\sqrt{2!}} \\ \frac{\delta^{[a_1}_{b}\,\ell^{a_2]}}{\sqrt{2!}} & 0 \end{pmatrix} .
\end{align}

\paragraph{Equations of motion:}
By using an identity,
\begin{align}
 \ell^b\wedge P_{ba} = \ell^b\wedge \bigl[\MA_{bac}\,\ell^c - *\bigl(\ell_{b}\wedge\ell_{a}\bigr)\bigr] = 2\,\cJ_a \,,
\end{align}
the equations of motion \eqref{eq:eom-En} can be expressed as
\begin{align}
 \ell^b\wedge \rmd P_{ba} &= -\frac{1}{2}\,f_{cd}{}^b\,\ell^c\wedge \ell^d \wedge P_{ba} 
\nn\\
 &\quad -\frac{1}{3}\,f_a{}^{bcd}\, \bigl[ \ell_b\wedge\ell_c\wedge\ell_d + \bigl(3\,\MA_{bce}\,H^e_d - \MA_{bcd}\bigr) *1 - \MA_{bcd}\,\MA_3\bigr] \,,
\label{eq:eom-En-2}
\end{align}
where we have used $\rmd\ell^b=-\frac{1}{2}\,f_{cd}{}^b\,\ell^c\wedge \ell^d$\,. 
We then consider a projection,
\begin{align}
 H_a^c\,\ell^b\wedge \rmd P_{bc} &= H_a^c\Bigl\{ -\frac{1}{2}\,f_{cd}{}^b\,\ell^c\wedge \ell^d \wedge P_{bc} 
\nn\\
 &\quad\qquad -\frac{1}{3}\,f_c{}^{bcd}\, \bigl[ \ell_b\wedge\ell_c\wedge\ell_d + \bigl(3\,\MA_{bce}\,H^e_d - \MA_{bcd}\bigr) *1 - \MA_{bcd}\,\MA_3\bigr]\Bigr\} \,.
\label{eq:eom-En-3}
\end{align}
When $n=3$, $H_a^c=\delta_a^c$ and they are equivalent to the equations of motion \eqref{eq:eom-En-2}, while when $n=4$, one equation has been projected out. 
In fact, as we show in Appendix \ref{app:eom}, Eq.~\eqref{eq:eom-En-3} is equivalent to
\begin{align}
 H_a^c\,\ell^b\wedge \Bigl(\rmd P_{bc} - \frac{1}{2}\,f_{[b}{}^{def}\,P_{c]d}\wedge P_{ef} + \frac{1}{2}\,f_{bc}{}^d\,P^e\wedge P_{de}\Bigr) = 0\,.
\label{eq:dP-projected}
\end{align}
To be more precise, when $n=3$ we can show the equivalence without any assumption, but when $n=4$ we need to assume $f_{ab}{}^a=0$\,. 
Then, the equations of motion \eqref{eq:dP-projected} imply
\begin{align}
 \rmd P_{ab} = \frac{1}{2}\,f_{[a}{}^{cde}\,P_{b]c}\wedge P_{de} - \frac{1}{2}\,f_{ab}{}^c\,P^d\wedge P_{cd} \,.
\label{eq:dP-1}
\end{align}
In fact, we can directly show that Eq.~\eqref{eq:dP-1} is identically satisfied in $n=3$ (see Appendix \ref{app:n=3}). 
Thus, the equations of motion in $n=3$ are automatically satisfied and the membrane is non-dynamical even in the curved background given in Eq.~\eqref{eq:En-geometry1}. 
In $n=4$, the projected equations \eqref{eq:dP-projected} are satisfied under the equations of motion, but they do not lead to Eq.~\eqref{eq:dP-1}. 

In fact, as we show in Appendix \ref{app:MC-En}, the relation \eqref{eq:dP-1}, which is suggested by the equations of motion, is equivalent to the MC equation of the $\cE_n$ algebra,
\begin{align}
 \rmd \cP^A = \frac{1}{2}\,\cF_{BC}{}^A\,\cP^B\wedge\cP^C \,,
\label{eq:MC-En}
\end{align}
where $\cF_{BC}{}^A$ are the structure constants of the $\cE_n$ algebra. 
Thus, in $n=3$, the generalized displacement $\cP^A$ satisfies the MC equation, which generalizes the Abelian one given in Eq.~\eqref{eq:MC-Abelian-U}. 
In $n=4$, $\cP^A$ does not satisfy the MC equation similar to the Abelian case, and we cannot perform the full $U$-duality transformation. 

\paragraph{Non-Abelian $U$-duality:}
In $n=3$, non-Abelian $U$-duality is realized as a redefinition of the $\cE_n$ generators,
\begin{align}
 T'_A = C_A{}^B\,T_B\,,
\label{eq:NAUD}
\end{align}
where $C_A{}^B$ is an element of the $U$-duality group $\SL(2)\times \SL(3)$. 
Under the redefinition, the structure constants are transformed as
\begin{align}
 \cF'_{AB}{}^C = C_A{}^D\,C_B{}^E\, \bigl(C^{-1}\bigr){}_F{}^C\,\cF_{DE}{}^F\,. 
\end{align}
In order to keep the MC 1-form $\cP$ invariant, the components should be transformed as
\begin{align}
 \cP'^A = \bigl(C^{-1}\bigr){}_B{}^A\,\cP^B\,.
\end{align}
Then, the MC equation \eqref{eq:MC-En} is manifestly covariant under non-Abelian $U$-duality \eqref{eq:NAUD}. 
The $\eta$-form is also transformed covariantly
\begin{align}
 \bm{\eta}'_{AB} = C_A{}^C\,C_B{}^D\,\bm{\eta}_{CD} \,,
\end{align}
and by further transforming the constant matrix as
\begin{align}
 \hat{\cH}'_{AB} = C_A{}^C\,C_B{}^D\,\hat{\cH}_{CD} \,,
\end{align}
the self-duality relation \eqref{eq:SD-NAUD} is also manifestly covariant under \eqref{eq:NAUD}. 

If a solution $x^i(\sigma)$ of membrane theory is given, we can explicitly compute the 1-form fields $\cP^A(\sigma)$. 
After the change of generators \eqref{eq:NAUD}, the 1-form fields are transformed as $\cP'^A = \bigl(C^{-1}\bigr){}_B{}^A\,\cP^B$\,.
We can also introduce a new group element $g'(x')=\Exp{x'^a\,T'_a}$, and through the relation \eqref{eq:cP-def}, we can compute the 1-form $P'^A=M'_B{}^A\, \cP'^B$ in the dual theory. 
Since the first component $P'^a$ has been identified as the left-invariant 1-form $\ell'^a$, by solving
\begin{align}
 P'^a\,T'_a = g'^{-1}(x')\,\rmd g'(x')\,,
\end{align}
we can in principle determine the dual solution $x'^i(\sigma)$\,.

\section{Discussion}
\label{sec:discussion}

In this paper, we have studied membrane theory in a curved background \eqref{eq:En-geometry1}, which has the symmetry of the $\cE_n$ algebra. 
Similar to the case of Abelian $U$-duality, we can show that the generalized displacement $\cP^A$ satisfies the MC equation of the $\cE_n$ algebra only when $n=3$. 
Both the MC equation and the self-duality relation for $\cP^A$ are manifestly covariant under non-Abelian $U$-duality \eqref{eq:NAUD} (which is a redefinition of the $\cE_n$ generators) and we have naturally extended the standard story of Abelian $U$-duality to the non-Abelian setup. 
In $n=4$, we face the difficulty already known in the Abelian case, and $\cP^A$ do not satisfy the MC equation even under the equations of motion. 

In addition to the membrane, M-theory contains the M5-brane as well (see \cite{1305.2258,1607.04265,1712.10316} for M5-brane theory in $U$-duality-covariant approaches). 
Again in the M5-brane theory, the equations of motion will not generally provide the MC equation. 
The only exceptional case will be $n=6$, where the M5-brane becomes space-filling. 
There, the generalized displacement is extended as $\cP^A=\bigl(\cP^a,\,\frac{\cP_{a_1a_2}}{\sqrt{2!}},\,\frac{\cP_{a_1\cdots a_5}}{\sqrt{5!}}\bigr)$ (see for example \cite{1712.10316}), and the number of the non-trivial components of the MC equations is $\frac{n!}{2!\,(n-2)!}+\frac{n!}{5!\,(n-5)!}$, corresponding to $\cP_{a_1a_2}$ and $\cP_{a_1\cdots a_5}$. 
The dynamical fields on the M5-brane are $x^i$ and the 2-form gauge field $A_{\WSa\WSb}$ ($\WSa,\WSb=0,\dotsc,5$) and, naively, the number of the equations of motion coincides with that of the non-trivial MC equations when $n=6$.\footnote{Here we have not taken into account of the self-duality relation for the gauge field.} 
Thus, we expect that $U$-duality symmetry in the M5-brane theory can be realized for $n=6$. 
For $n>6$, the number of the equations of motion is smaller and the full MC equation will not be reproduced. 
In order to examine this possibility, it is important to construct the $\cE_n$ algebra for $n=6$ or higher. 

As it is well-known, when the (self-dual) field strength on the M5-brane is non-vanishing, the M2-brane is induced on the M5-brane. 
Then, the dynamics of the induced M2-brane will be described by the 2-form gauge fields $A_{\WSa\WSb}$ on the M5-brane. 
For example, in section 6 of \cite{1308.2231}, the gauge fields are dualized to the embedding functions $x^i$ of the M2-brane, and the membrane action has been reproduced from the M5-brane action. 
Then, it is interesting to consider the following possibility. 
As we discussed in this paper, in $n=6$, membrane theory does not have the $E_6$ $U$-duality symmetry. 
However, if the $E_6$ $U$-duality symmetry is realized in the topological (or space-filling) M5-brane theory, it is interesting to interpret the topological M5-brane theory as the $E_6$-covariant membrane theory. 
Since the M5-brane is space-filling $x^i$ will be non-dynamical, and only the gauge fields $A_{\WSa\WSb}$ are dynamical, which describe the fluctuation of the membrane. 
If $\cP^A$ satisfies the MC equation, we can perform non-Abelian $U$-duality. 
This approach may resolve the issue of $U$-duality in membrane theory. 
Moreover, in the approach of \cite{1607.04265,1712.10316}, gauge fields on the worldvolume are introduced as the diffeomorphism parameters along the dual direction in the extended spacetime. 
In other words, the gauge fields are interpreted as the fluctuation along the dual directions in the extended spacetime. 
Since the number of diffeomorphism parameters along the dual direction is always the same as the number of the non-trivial components of the MC equations, naively we can expect that the MC equation is realized under the equations of motion even for higher $n$. 
For example, in $n=8$, it will be impossible to realize the $E_8$ duality symmetry in M5-brane theory. 
However, there, the Kaluza--Klein monopole (KKM) is space-filling, and its worldvolume theory may have the $E_8$ $U$-duality symmetry. 
If so, it may be possible to regard the topological KKM theory as the $E_8$ M5-brane theory. 
We hope to work on this in the future. 

\subsection*{Acknowledgments}

The work by Y.S.\ is supported by JSPS Grant-in-Aids for Scientific Research (C) 18K13540 and (B) 18H01214.

\subsection*{Note added}

To clarify the connection with the $\cE$-model \cite{1508.05832}, here we show the classical current algebra.\footnote{We would like to thank an anonymous referee for the suggestion.} 
One of the defining properties of the $\cE$-model is the current algebra
\begin{align}
 \{j_A(\sigma),\,j_B(\sigma')\} = \cF_{AB}{}^C\,j_C(\sigma)\,\delta(\sigma-\sigma') + \eta_{AB}\,\partial_\sigma \delta(\sigma-\sigma) \,. 
\label{eq:E-model}
\end{align}
In the context of the PL $T$-duality, the spatial component of $\cJ_A \equiv \hat{\cH}_{AB} * \cP^B$ plays the role of the current
\begin{align}
 j_A(\sigma) = \cJ_{A\sigma} = E_A{}^M\bigl(x(\sigma)\bigr)\,Z_M(\sigma)\,,\qquad 
 Z_M(\sigma) \equiv (\cH_{MN}*P^N)_{\sigma} = \begin{pmatrix} p_m \\ \partial_\sigma x^m \end{pmatrix}.
\label{eq:jA-def}
\end{align}
Here, $p_m \equiv -g_{mn}\,\sqrt{-h}\,h^{0\alpha}\,\partial_\alpha x^n + B_{mn}\, \epsilon^{0\alpha}\,\partial_\alpha x^n$ is the canonical conjugate momenta of $x^m(\sigma)$ in the Hamiltonian formulation. 
As was shown in \cite{hep-th:9305073,hep-th:9710163}, this current satisfies the algebra \eqref{eq:E-model} by means of the equal-time Poisson-bracket $\{x^m(\sigma),\,p_n(\sigma')\}=\delta^m_n\,\delta(\sigma-\sigma')$\,. 

We here provide a brief sketch of the extension of the current algebra to the case of the membrane theory. 
For this purpose, we employ the analysis of Ref.~\cite{1208.1232}, where a canonical analysis of the membrane theory in a flat space was worked in the $\SL(5)$ covariant manner. 
There, the generalized momenta $Z_M(\sigma)$ defined in Eq.~\eqref{eq:jA-def} are extended to $Z_I(\sigma)$ which are defined as the spatial components of $\cJ_I$ defined in Eq.~\eqref{eq:cJ-Abelian-U}. 
Namely, $Z_I(\sigma)$ is given by
\begin{align}
 Z_I(\sigma) &\equiv \frac{1}{2}\,\epsilon^{0\alpha\beta}\, \cJ_{I\alpha\beta} = \begin{pmatrix} p_i \\ \frac{\epsilon^{0\alpha\beta}\,\partial_{\alpha} x^{[i_1} \,\partial_{\beta} x^{i_2]}}{\sqrt{2!}} \end{pmatrix} ,
\\
 p_i &\equiv - g_{ij}\,\sqrt{-h}\,h^{0\alpha} \,\partial_\alpha x^j - \frac{1}{2}\,\epsilon^{0\alpha\beta}\,\MA_{ijk}\,\partial_{\alpha} x^j \,\partial_{\beta} x^k\,.
\end{align}
Then the equal-time Poisson-bracket $\{x^i(\sigma),\,p_j(\sigma')\}=\delta^i_j\,\delta^2(\sigma-\sigma')$ leads to
\begin{align}
 \{Z_I(\sigma),\,Z_J(\sigma')\} = \rho^\alpha_{IJ}(\sigma)\,\partial_{\alpha} \delta^2(\sigma-\sigma')\,,\qquad
 \rho^\alpha_{IJ}(\sigma) \equiv \epsilon^{0\alpha\beta}\, \eta_{IJ;k}\,\partial_\beta x^k(\sigma)\,.
\end{align}
In the case of our interest ($n\leq 4$), defining the current $j_A(\sigma)$ as
\begin{align}
 j_A(\sigma) \equiv E_A{}^I\bigl(x^i(\sigma)\bigr)\,Z_I(\sigma)\,,
\end{align}
we obtain a natural extension of the current algebra \eqref{eq:E-model},\footnote{Here we have used the following identity \cite{1208.1232} for $n\leq 4$ and arbitrary vectors $\Lambda_1^I$ and $\Lambda_2^I$ (the arbitrary parameter $K$ in \cite{1208.1232} is chosen as $K=1$):
\begin{align*}
 \bigl\{\Lambda_1^I\bigl(x^i(\sigma)\bigr)\,Z_I(\sigma),\,\Lambda_2^J\bigl(x^i(\sigma')\bigr)\,Z_J(\sigma')\bigr\}
 = - (\gLie_{\Lambda_1} \Lambda_2)^I\bigl(x^i(\sigma)\bigr)\,Z_I(\sigma)\,\delta^2(\sigma-\sigma') 
 - \bigl(\rho^\alpha_{IJ}\,\Lambda_1^I\,\Lambda_2^J \bigr)(\sigma')\,\partial_\alpha \delta^2(\sigma-\sigma')\,.
\end{align*}}
\begin{align}
 \bigl\{j_A(\sigma),\,j_B(\sigma')\bigr\} = \cF_{AB}{}^C\,j_C(\sigma)\,\delta^2(\sigma-\sigma') 
 - \bigl(\rho^\alpha_{IJ}\,E_A^I\,E_B^J \bigr)(\sigma')\,\partial_\alpha \delta^2(\sigma-\sigma')\,,
\label{eq:M2-current}
\end{align}
where the coefficient in the second term is no longer constant.
It would be interesting to study the membrane extension of the $\cE$-model that is defined by the current algebra \eqref{eq:M2-current} and the Hamiltonian $H \equiv \frac{1}{2}\int\rmd^2\sigma\, \hat{\cH}^{AB}\,j_A(\sigma)\,j_B(\sigma)$\,. 

\appendix

\section{Formulas in $n=4$}
\label{app:n=4}

In $n=4$, relations between the non-geometric fields and the standard fields are given as
\begin{align}
\begin{split}
 &\tilde{g}_{ij} = K^{\frac{1}{3}}\,E_{ij}\,,\qquad 
 E_{ij} \equiv g_{ij} + \frac{1}{2}\,\MA_{i}{}^{kl}\,\MA_{klj}\,, \qquad 
 \Omega^{ijk} = - E^{il}\,g^{jm}\,g^{kn}\,\MA_{lmn} = -K\,\MA^{ijk}\,,
\\
 &\Omega_{ijk}= -K^{-1}\,\MA_{ijk}\,,\qquad \det \bigl(E_{ij}\bigr) = K^{-3}\,\det \bigl(g_{ij}\bigr)\,,\qquad \sqrt{-\tilde{g}} = K^{-\frac{5}{6}} \sqrt{-g}\,,
\end{split}
\end{align}
where the indices of $\MA_{ijk}$ and $\Omega^{ijk}$ are raised or lowered with the metric $g_{ij}$ and $\tilde{g}_{ij}$, respectively. 
In $n=4$, it is useful to parameterize $\MA_{ijk}$ and $\Omega^{ijk}$ as
\begin{align}
 \MA_{ijk} = \varepsilon_{ijkl}\,\MA^l\,,\qquad 
 \Omega^{ijk} = \varepsilon^{ijkl}\,\Omega_l\,,
\end{align}
and then we find the following relations:
\begin{align}
\begin{split}
 &\Omega^i = -K^{-\frac{1}{6}}\,\MA^i \,,\qquad \Omega_i = -K^{\frac{1}{6}}\,\MA_i \,,\qquad 
  K=\frac{1}{1-g_{ij}\,\MA^i\,\MA^j}=\frac{1}{1-\tilde{g}^{ij}\,\Omega_i\,\Omega_j}\,,
\\
 &g_{ij} = K^{\frac{2}{3}}\bigl(\tilde{g}_{ij}-\Omega_i\,\Omega_j\bigr)\,,\qquad 
 \tilde{g}_{ij} = K^{\frac{1}{3}}\bigl(g_{ij}-\MA_i\,\MA_j\bigr)\,.
\end{split}
\end{align}

\section{Results specific to $n=3$}
\label{app:n=3}

In this Appendix, by considering $n=3$, we show Eq.~\eqref{eq:dP-1}, namely,
\begin{align}
 \rmd P_{ab} = \frac{1}{2}\,f_{[a}{}^{cde}\,P_{b]c}\wedge P_{de} - \frac{1}{2}\,f_{ab}{}^c\,P^d\wedge P_{cd} \,.
\end{align}

Using the dual metric $\tilde{g}_{ij}$ defined in Eq.~\eqref{eq:dual-fields}, we define the anti-symmetric tensor
\begin{align}
 \tilde{\varepsilon}_{ijk} \equiv \sqrt{\abs{\tilde{g}}}\,\epsilon_{ijk}\,,\qquad 
 \tilde{\varepsilon}^{ijk} \equiv \frac{1}{\sqrt{\abs{\tilde{g}}}}\,\epsilon^{ijk}\,,\qquad 
 \epsilon^{012} \equiv 1\equiv -\epsilon_{012}\,.
\end{align}
In $n=3$, the tri-vector $\Omega^{ijk}$ should be proportional to $\tilde{\varepsilon}^{ijk}$ and we define
\begin{align}
 \Omega^{ijk} \equiv \omega \,\tilde{\varepsilon}^{ijk} \,.
\end{align}
From Eq.~\eqref{eq:En-geometry2}, we obtain the standard supergravity fields as
\begin{align}
 g_{ij} = \frac{1}{(1-\omega^2)^{2/3}}\, \tilde{g}_{ij} \,, \qquad 
 \MA_{ijk} = -\frac{\omega}{1-\omega^2}\,\tilde{\varepsilon}_{ijk} = -\omega\,\varepsilon_{ijk} \,,
\end{align}
where we have defined
\begin{align}
 \varepsilon_{ijk} \equiv \sqrt{\abs{g}}\,\epsilon_{ijk} = \frac{1}{1-\omega^2}\,\tilde{\varepsilon}_{ijk}\,,\qquad 
 \varepsilon^{ijk} \equiv \frac{1}{\sqrt{\abs{g}}}\,\epsilon^{ijk} = \bigl(1-\omega^2\bigr)\,\tilde{\varepsilon}^{ijk} \,.
\end{align}
Using $\Lie_{v_a}\tilde{\varepsilon}^{ijk}=0$ (which follows from $\Lie_{v_a}\tilde{g}_{ij}=0$), Eq.~\eqref{eq:Lie-dual} reduces to
\begin{align}
 \Lie_{v_a} \omega = - \frac{1}{3!}\,f_a{}^{bcd}\,\tilde{\varepsilon}_{bcd} \quad\Leftrightarrow\quad \rmd \omega = - \frac{1}{3!}\,f_a{}^{bcd}\,\tilde{\varepsilon}_{bcd}\,\ell^a\,.
\end{align}

Another important relation specific to $n=3$ is
\begin{align}
 \varepsilon^{ab}{}_c\,\ell^c = *\bigl(\ell^a\wedge\ell^b\bigr) \,.
\end{align}
This allows us to simplify $P_{ab}$ as
\begin{align}
 P_{ab} \equiv \MA_{abc}\,\ell^c - *\bigl(\ell_a\wedge\ell_b\bigr) = -\frac{\omega}{1-\omega^2}\,\tilde{\varepsilon}_{abc}\,\ell^c - \frac{1}{1-\omega^2}\, \tilde{\varepsilon}_{abc}\,\ell^c
 = -\frac{1}{1-\omega}\,\tilde{\varepsilon}_{abc}\,\ell^c \,,
\end{align}
and we obtain
\begin{align}
 \rmd P_{ab} = -\frac{\rmd \omega}{(1-\omega)^2}\,\tilde{\varepsilon}_{abc}\wedge\ell^c
 - \frac{\rmd\ln\sqrt{\abs{\tilde{g}}}}{1-\omega}\,\tilde{\varepsilon}_{abc}\wedge\ell^c 
 - \frac{1}{1-\omega}\,\tilde{\varepsilon}_{abc}\,\rmd\ell^c\,.
\end{align}
Now, let us rewrite each term on the right-hand side. 
The first term is
\begin{align}
 -\frac{\rmd \omega}{(1-\omega)^2}\,\tilde{\varepsilon}_{abc}\wedge\ell^c
 &= \frac{\tilde{\varepsilon}_{abc}}{3!\,(1-\omega)^2}\, f_e{}^{d_1d_2d_3}\,\tilde{\varepsilon}_{d_1d_2d_3}\,\ell^e\wedge \ell^c 
 = -\frac{1}{3!\,(1-\omega)}\, f_e{}^{d_1d_2d_3}\,\tilde{\varepsilon}_{d_1d_2d_3}\,\ell^e\wedge P_{ab} 
\nn\\
 &= -\frac{1}{12}\, f_{e_1}{}^{d_1d_2d_3}\,\tilde{\varepsilon}_{d_1d_2d_3}\,\tilde{\varepsilon}^{e_1e_2e_3}\,P_{e_2e_3} \wedge P_{ab} 
\nn\\
 &= -\frac{1}{12}\, f_{[a|}{}^{d_1d_2d_3}\,\tilde{\varepsilon}_{d_1d_2d_3}\,\tilde{\varepsilon}^{e_1e_2e_3}\,P_{e_1e_2} \wedge P_{e_3|b]} 
\nn\\
 &= \frac{1}{2}\, f_{[a}{}^{c_1c_2c_3}\,P_{b]c_1} \wedge P_{c_2c_3}\,,
\end{align}
where we have used the Schouten identity, $f_{[e_1}{}^{d_1d_2d_3}\,P_{e_2e_3} \wedge P_{a]b}=0$. 
The second term is
\begin{align}
 -\frac{\rmd\ln\sqrt{\abs{\tilde{g}}}}{1-\omega}\,\tilde{\varepsilon}_{abc}\wedge\ell^c 
 = \frac{f_{de}{}^d}{1-\omega}\,\tilde{\varepsilon}_{abc}\,\ell^e\wedge\ell^c \,,
\end{align}
and the third term is
\begin{align}
 -\frac{1}{1-\omega}\,\tilde{\varepsilon}_{abc}\,\rmd\ell^c
 &= \frac{1}{2}\,\frac{1}{1-\omega}\,\tilde{\varepsilon}_{abc}\,f_{de}{}^c\,\ell^d\wedge\ell^e
 = \frac{3}{2}\,\frac{1}{1-\omega}\,\tilde{\varepsilon}_{d[ab}\,f_{c]e}{}^c\,\ell^d\wedge\ell^e
\nn\\
 &= \frac{3}{2}\,\frac{1}{1-\omega}\,\tilde{\varepsilon}_{de[a}\,f_{bc]}{}^c\,\ell^d\wedge\ell^e
\nn\\
 &= -\frac{f_{de}{}^d}{1-\omega}\,\tilde{\varepsilon}_{abc}\,\ell^e\wedge\ell^c - \frac{1}{2}\,f_{ab}{}^c\,\ell^d\wedge P_{cd}\,.
\end{align}
Thus, we obtain
\begin{align}
 \rmd P_{ab} = \frac{1}{2}\, f_{[a}{}^{c_1c_2c_3}\,P_{b]c_1} \wedge P_{c_2c_3} - \frac{1}{2}\,f_{ab}{}^c\,P^d\wedge P_{cd}\,,
\end{align}
which is an identity in $n=3$\,. 

\section{Maurer--Cartan equation for the $\cE_n$ algebra}
\label{app:MC-En}

In this appendix, for $n\leq 4$, we show that 
\begin{align}
 \rmd P_{ab} = \frac{1}{2}\,f_{[a}{}^{cde}\,P_{b]c}\wedge P_{de} - \frac{1}{2}\,f_{ab}{}^c\,P^d\wedge P_{cd} \,,
\label{eq:dP-1A}
\end{align}
given in Eq.~\eqref{eq:dP-1} is equivalent to the $U$-duality-manifest MC equation 
\begin{align}
 \rmd \cP^A - \frac{1}{2}\,\cF_{BC}{}^A\,\cP^B\wedge\cP^C = 0 \,,
\label{eq:MC-En-A}
\end{align}
where
\begin{align}
 \bigl(\cP^A\bigr) \equiv \begin{pmatrix} \cP^a \\ \frac{\cP_{a_1a_2}}{\sqrt{2!}}
 \end{pmatrix} \equiv \begin{pmatrix} r^a + \frac{1}{2}\,\Pi^{ab_1b_2}\,\cP_{b_1b_2} \\ \frac{a_{a_1}{}^{b_1}\,a_{a_2}{}^{b_2}\,P_{b_1b_2}}{\sqrt{2!}}
 \end{pmatrix}.
\end{align}

By using the explicit form of $\cF_{BC}{}^A$, The MC equation \eqref{eq:MC-En-A} is equivalent to
\begin{align}
 \rmd \cP^a &= \frac{1}{2}\,f_{bc}{}^a\,\cP^b\wedge\cP^c + \frac{1}{2}\,f_b{}^{c_1c_2a}\,\cP^b\wedge\cP_{c_1c_2}\,,
\label{eq:MC-1}
\\
 \rmd \cP_{a_1a_2} &= 2\,f_{b[a_1}{}^c\,\cP^b\wedge\cP_{a_2]c} + \frac{1}{2}\,f_{a_1a_2}{}^b\,\cP^c\wedge\cP_{cb} 
 + \frac{1}{2}\,f_{[a_1}{}^{b_1b_2c}\,\cP_{a_2]c}\wedge \cP_{b_1b_2} \,.
\label{eq:MC-2}
\end{align}
As we show later, the former follows from the latter. 
Thus, in the following, we show that the latter is equivalent to Eq.~\eqref{eq:dP-1A}. 
To this end, we employ the following identities \cite{1911.06320}:
\begin{align}
 &\rmd a_a{}^b = a_a{}^d \,f_{de}{}^b\, \ell^e \,,
\label{eq:da}
\\
 &\rmd \Pi^{a_1a_2a_3} = f_b{}^{a_1a_2a_3}\,r^b + 3\, f_{bc}{}^{[a_1}\,\Pi^{a_2a_3]c}\,r^b \,,
\label{eq:dPi}
\\
 &\bigl(a^{-1}\bigr){}_a{}^e\,\bigl(a^{-1}\bigr){}_b{}^f\,a_g{}^c\,f_{ef}{}^g = f_{ab}{}^c\,,
\label{eq:aaaf}
\\
 &a_a{}^e\, \bigl(a^{-1}\bigr){}_{f_1}{}^{b_1}\,\bigl(a^{-1}\bigr){}_{f_2}{}^{b_2}\,\bigl(a^{-1}\bigr){}_{f_3}{}^{b_3}\, f_e{}^{f_1f_2f_3} = f_a{}^{b_1b_2b_3} + 3\, f_{ac}{}^{[b_1}\,\Pi^{b_2b_3]c} \,,
\label{eq:aaaaft}
\\
 &3\,\bigl(f_{e[c}{}^{a_1}\,\delta_{d]}^{[a_2}\,\Pi^{b_1b_2]e}
 - f_{e[c}{}^{a_2}\,\delta_{d]}^{[a_1}\,\Pi^{b_1b_2]e} \bigr)
 + f_{cd}{}^{[a_1}\,\Pi^{a_2]b_1b_2} = 0 \,,
\label{eq:id1}
\\
 &f_d{}^{b_1b_2c}\,\Pi^{a_1a_2d}
 -3\,f_d{}^{a_1a_2[b_1}\,\Pi^{b_2c]d}
 = 3\,f_{de}{}^{[c}\,\Pi^{b_1b_2]d}\,\Pi^{a_1a_2e}
 - 4\,f_{de}{}^{[a_1}\,\Pi^{a_2]d[b_1}\,\Pi^{b_2]ec} \,,
\label{eq:id2}
\\
 &f_{ab}{}^c\,\Pi^{abd} =0 \,. 
\label{eq:id3}
\end{align}
By using Eqs.~\eqref{eq:da}, \eqref{eq:aaaf}, and \eqref{eq:aaaaft}, Eq.~\eqref{eq:MC-2} becomes
\begin{align}
 \rmd P_{a_1a_2}
 &= - 2\,f_{b[a_1}{}^c\,\ell^{b}\wedge P_{a_2]c} 
   + 2\,f_{b[a_1}{}^c\,\tilde{\cP}^{b}\wedge P_{a_2]c} 
   + \frac{1}{2}\,f_{a_1a_2}{}^b\,\tilde{\cP}^c\wedge P_{cb}
\nn\\
 &\quad
   + \frac{1}{2}\,f_{[a_1}{}^{c_1c_2d}\,P_{a_2]d}\wedge P_{c_1c_2}
   + \frac{3}{2}\,f_{e[a_1}{}^{[c_1}\,\tilde{\Pi}^{c_2d]e}\,P_{a_2]d}\wedge P_{c_1c_2}
\nn\\
 &= \frac{1}{2}\,f_{[a_1}{}^{cde}\,P_{a_2]c}\wedge P_{de} - \frac{1}{2}\,f_{a_1a_2}{}^b\,P^c\wedge P_{bc}
\nn\\
 &\quad + \frac{1}{4}\,\Bigl(6\,f_{e[a_1}{}^{d}\,\delta_{a_2]}^{[b} \,\tilde{\Pi}^{c_1c_2]e}
 + f_{a_1a_2}{}^d\,\tilde{\Pi}^{bc_1c_2}\Bigr)\,P_{db}\wedge P_{c_1c_2}\,,
\label{eq:dP-vanish}
\end{align}
where we have defined
\begin{align}
 \tilde{\cP}^a \equiv a_b{}^a\,\cP^b\,,\qquad \tilde{\Pi}^{abc} \equiv a_d{}^a\, a_e{}^b\, a_f{}^c\,\Pi^{def}\,.
\end{align}
By further using Eq.~\eqref{eq:id1}, we can show that the last line of Eq.~\eqref{eq:dP-vanish} vanishes.
Then, we have shown that Eq.~\eqref{eq:MC-2} is equivalent to Eq.~\eqref{eq:dP-1A}. 

In the remainder of this Appendix, we show that Eq.~\eqref{eq:MC-1} is trivially satisfied under Eq.~\eqref{eq:MC-2}. 
By using the explicit form of $\cP^a$, the left-hand side of Eq.~\eqref{eq:MC-1} is
\begin{align}
 \rmd \cP^a = \frac{1}{2}\,f_{bc}{}^a\,r^b\wedge r^c + \frac{1}{2}\,f_c{}^{b_1b_2a}\,r^c\wedge P_{b_1b_2} + \frac{3}{2}\,r^c\,f_{cd}{}^{[b_1}\,\Pi^{b_2a]d}\,P_{b_1b_2} + \frac{1}{2}\,\Pi^{b_1b_2a}\,\rmd \cP_{b_1b_2} \,.
\end{align}
Then, by using Eqs.~\eqref{eq:MC-2} and \eqref{eq:id3}, Eq.~\eqref{eq:MC-1} is equivalent to
\begin{align}
\begin{split}
 &\frac{1}{2}\,\Pi^{a_1a_2a}\,f_{ba_1}{}^c\,\Pi^{bd_1d_2}\,P_{d_1d_2}\wedge P_{a_2c}
 + \frac{1}{4}\,f_b{}^{c_1c_2a}\,\Pi^{be_1e_2}\,P_{c_1c_2}\wedge P_{e_1e_2}
\\
 &- \frac{1}{4}\,f_{a_1}{}^{b_1b_2c}\,\Pi^{a_2aa_1}\,P_{ca_2}\wedge P_{b_1b_2}
 - \frac{1}{8}\,f_{bc}{}^a\,\Pi^{be_1e_2}\,\Pi_{cf_1f_2}\,P_{e_1e_2}\wedge P_{f_1f_2} = 0\,.
\end{split}
\label{eq:dPa-1}
\end{align}
By further using an identity
\begin{align}
 & \frac{1}{4}\,f_b{}^{c_1c_2a}\,\Pi^{be_1e_2}\,P_{c_1c_2}\wedge P_{e_1e_2}
  - \frac{1}{4}\,f_{a_1}{}^{b_1b_2c}\,\Pi^{a_2aa_1}\,P_{ca_2}\wedge P_{b_1b_2}
  - \frac{1}{8}\,f_{bc}{}^a\,\Pi^{be_1e_2}\,\Pi^{cf_1f_2}\,P_{e_1e_2}\wedge P_{f_1f_2} 
\nn\\
 &+ \frac{1}{4}\,f_{ed}{}^{b_2}\,\Pi^{db_1a}\,\Pi^{a_1a_2e}\,P_{a_1a_2}\wedge P_{b_1b_2}
  + \frac{1}{2}\,f_{de}{}^{a_1}\,\Pi^{a_2db_1}\,\Pi^{b_2ea}\,P_{b_1b_2}\wedge P_{a_1a_2}=0\,,
\end{align}
which follows from Eq.~\eqref{eq:id2}, we can show that Eq.~\eqref{eq:dPa-1} is equivalent to
\begin{align}
 \frac{1}{4}\,f_{ed}{}^{b_2}\,\Pi^{db_1c}\,\Pi^{a_1a_2e}\,P_{b_1b_2}\wedge P_{da_1}
 -\frac{1}{2}\,f_{de}{}^{a_1}\,\Pi^{a_2db_1}\,\Pi^{b_2ec}\,P_{b_1b_2}\wedge P_{a_1a_2} = 0\,.
\end{align}
We can easily see that this equality follows from Eq.~\eqref{eq:id1}. 
Thus, Eq.~\eqref{eq:MC-1} is always satisfied when Eq.~\eqref{eq:MC-2} is satisfied. 

\section{Rewriting the equations of motion}
\label{app:eom}

Here, we rewrite the equations of motion Eq.~\eqref{eq:eom-En-2},
\begin{align}
 \ell^b\wedge \rmd P_{ba} &= -\frac{1}{2}\,f_{cd}{}^b\,\ell^c\wedge \ell^d \wedge P_{ba} 
\nn\\
 &\quad -\frac{1}{3}\,f_a{}^{bcd}\, \bigl[ \ell_b\wedge\ell_c\wedge\ell_d + \bigl(3\,\MA_{bce}\,H^e_d - \MA_{bcd}\bigr) *1 - \MA_{bcd}\,\MA_3\bigr] \,,
\end{align}
into a more convenient form. 
Only when $n=4$, we assume that $f_{ab}{}^a=0$\,. 

We begin by rewriting the first term on the right-hand side as
\begin{align}
 -\frac{1}{2}\,f_{cd}{}^b\,\ell^c\wedge \ell^d \wedge P_{ba} = 
 -\frac{1}{2}\,f_{cd}{}^b\,\ell^c\wedge \ell^d \wedge \bigl[\MA_{bae}\,\ell^e + *\bigl(\ell_a\wedge\ell_b\bigr)\bigr]\,.
\end{align}
Here, by using Eq.~\eqref{eq:invariant-metric}, the second term vanishes 
\begin{align}
 f_{cd}{}^b\,\ell^c\wedge \ell^d \wedge *\bigl(\ell_a\wedge\ell_b\bigr) 
 = 2\,H^c_a\,f_{c(d}{}^b\,g_{e)b}\, H^{de} *1 = 0\,.
\end{align}
Then, in $n=3$, by using the Schouten identity $A_{[cdae]}=0$, we obtain
\begin{align}
 -\frac{1}{2}\,f_{cd}{}^b\,\ell^c\wedge \ell^d \wedge P_{ba} 
 = \frac{1}{2}\,f_{ac}{}^b\,\MA_{bde}\,\ell^c\wedge \ell^d \wedge \ell^e \,.
\end{align}
In $n=4$, by using the Schouten identity $A_{[cdbae]}=0$, we obtain the same relation
\begin{align}
 -\frac{1}{2}\,f_{cd}{}^b\,\ell^c\wedge \ell^d \wedge P_{ba} 
 = \frac{1}{2}\,f_{ac}{}^b\,\MA_{bde}\,\ell^c\wedge \ell^d \wedge \ell^e \,,
\end{align}
although the assumption $f_{ab}{}^a=0$ has been used. 
Then, both in $n=3$ and $n=4$, we obtain
\begin{align}
 -\frac{1}{2}\,f_{cd}{}^b\,\ell^c\wedge \ell^d \wedge P_{ba}
 = \frac{1}{2}\,f_{ac}{}^b\,\MA_{bde}\,\ell^c\wedge \ell^d \wedge \ell^e
 = \frac{1}{2}\,f_{ac}{}^b\,\ell^c\wedge \ell^d \wedge P_{bd} \,,
\end{align}
where we have used $f_{ac}{}^b\,g_{bd}\,\ell^c\wedge * \ell^d = 0$\,.
The equations of motion then become
\begin{align}
 \ell^b\wedge \rmd P_{ba} &= \ell^c\wedge \Bigl(-\frac{1}{2}\,f_{ca}{}^b\,\ell^d \wedge P_{bd} \Bigr)
\nn\\
 &\quad -\frac{1}{3}\,f_a{}^{bcd}\, \bigl[ \ell_b\wedge\ell_c\wedge\ell_d + \bigl(3\,\MA_{bce}\,H^e_d - \MA_{bcd}\bigr) *1 - \MA_{bcd}\,\MA_3\bigr] \,.
\label{eq:eom-2}
\end{align}

Now, we rewrite the second line of Eq.~\eqref{eq:eom-2}.
In $n=3$, we can easily rewrite it as
\begin{align}
 -\frac{1}{3}\,f_a{}^{bcd}\, \bigl[ \ell_b\wedge\ell_c\wedge\ell_d + 2\,\MA_{bcd} *1 - \MA_{bcd}\,\MA_3\bigr]
 = \frac{1}{3}\,f_a{}^{bcd}\, \bigl(1 + 2\,\omega + \omega^2\bigr)\,\varepsilon_{bcd}\,*1\,. 
\end{align}
Using an identity $P_{ab} = -(1+\omega)\,\varepsilon_{abc}\,\ell^c$ in $n=3$, we also find
\begin{align}
 \ell^b\wedge \frac{1}{2}\,f_{[b}{}^{def}\,P_{a]d}\wedge P_{ef} 
 = (1+\omega)^2 \,f_{[b}{}^{dbe}\,\varepsilon_{a]de}*1 
 = \frac{1}{3}\,(1+\omega)^2\,f_{a}{}^{bcd}\,\varepsilon_{bcd}*1 \,,
\end{align}
and these show the following relation:
\begin{align}
 -\frac{1}{3}\,f_a{}^{bcd}\, \bigl[ \ell_b\wedge\ell_c\wedge\ell_d + 2\,\MA_{bcd} *1 - \MA_{bcd}\,\MA_3\bigr]
 = \ell^b\wedge \Bigl(\frac{1}{2}\,f_{[b}{}^{def}\,P_{a]d}\wedge P_{ef}\Bigr) \,. 
\end{align}
Then, the equations of motion \eqref{eq:eom-2} become
\begin{align}
 \ell^b\wedge \Bigl(\rmd P_{ba} - \frac{1}{2}\,f_{[b}{}^{def}\,P_{a]d}\wedge P_{ef} + \frac{1}{2}\,f_{ba}{}^d\,P^e\wedge P_{de}\Bigr) = 0\,.
\end{align}
In $n=4$, as we show below, we obtain a projected relation,
\begin{align}
 &H_a^c\,\Bigl(-\frac{1}{3}\,f_c{}^{bcd}\, \bigl[ \ell_b\wedge\ell_c\wedge\ell_d + \bigl(3\,\MA_{bce}\,H^e_d - \MA_{bcd}\bigr) *1 - \MA_{bcd}\,\MA_3\bigr]\Bigr) 
\nn\\
 &= H_a^c\,\ell^b\wedge \Bigl(\frac{1}{2}\,f_{[b}{}^{def}\,P_{c]d}\wedge P_{ef}\Bigr)\,.
\label{eq:id-n=4}
\end{align}
Then, by combining Eqs.~\eqref{eq:eom-2} and \eqref{eq:id-n=4}, we obtain
\begin{align}
 H_a^c\,\ell^b\wedge \Bigl(\rmd P_{bc} - \frac{1}{2}\,f_{[b}{}^{def}\,P_{c]d}\wedge P_{ef} + \frac{1}{2}\,f_{bc}{}^d\,P^e\wedge P_{de}\Bigr) = 0\,.
\label{eq:eom-3}
\end{align}
In $n=3$, this is equivalent to the equations of motion \eqref{eq:eom-En-2} while in $n=4$ this is equivalent to the projected components of the equations of motion. 

\paragraph{Derivation of Eq.~\eqref{eq:id-n=4}:}
In the following, we use several relations specific to $n=4$ (see Appendix \ref{app:n=4}) and also use the Schouten identity, $A_{[a_1\cdots a_5]}=0$. 

Let us begin by considering the following expansion:
\begin{align}
 &\frac{1}{2}\,f_{[a}{}^{efg}\,\ell^a\wedge P_{b]e}\wedge P_{fg}
\nn\\
 &= \underbrace{\frac{1}{2}\,f_{[a}{}^{efg}\,\ell^a\wedge *\bigl(\ell_{b]}\wedge\ell_e\bigr)\wedge * \bigl(\ell_f\wedge\ell_g\bigr)}_{\text{(A)}}
    +\underbrace{\frac{1}{2}\,f_{[a}{}^{efg}\,\MA_{b]ep}\,\MA_{fgq}\,\ell^a\wedge\ell^p\wedge\ell^q}_{\text{(B)}}
\nn\\
 &\quad \underbrace{-\frac{1}{2}\,f_{[a|}{}^{efg}\,\bigl[\MA_{|b]ep}\,\ell^a\wedge\ell^p\wedge *\bigl(\ell_f\wedge\ell_g\bigr) 
    + \MA_{fgq}\,\ell^a\wedge *\bigl(\ell_{|b]}\wedge\ell_e\bigr) \wedge\ell^q\bigr]}_{\text{(C)}}\,.
\end{align}
We rewrite each term as follows. 
The first term is
\begin{align}
 \text{(A)} &= \frac{1}{2}\,\bigl(f_a{}^{efg}\,H_f^a\,\ell_b^{\gamma_1}\,\ell_e^{\gamma_2}\,\ell_g^\beta\,\varepsilon_{\gamma_1\gamma_2\beta} 
 + f_b{}^{efg}\,\varepsilon_{\alpha\beta\gamma}\,\ell_e^\alpha\,\ell_f^\beta\,\ell_g^\gamma \bigr) *1
\nn\\
 &= \Bigl(-\frac{1}{3!}\,f_a{}^{efg}\,H_b^a\,\ell_e^{\gamma_1}\,\ell_f^{\gamma_2}\,\ell_g^\beta\,\varepsilon_{\gamma_1\gamma_2\beta} 
 + \frac{1}{2}\,f_b{}^{efg}\,\varepsilon_{\alpha\beta\gamma}\,\ell_e^\alpha\,\ell_f^\beta\,\ell_g^\gamma \Bigr) *1
\nn\\
 &= -\frac{1}{3}\,f_b{}^{efg}\, \ell_e\wedge \ell_f\wedge \ell_g 
 + \frac{1}{3!}\,f_a{}^{efg}\,N_b^a\,\ell_e^{\alpha}\,\ell_f^{\beta}\,\ell_g^\gamma\,\varepsilon_{\alpha\beta\gamma} *1 \,,
\end{align}
where we have defined $N_a^b\equiv \delta_a^b -H_a^b$, which is a projector along the orthogonal directions of the membrane.
The second term is
\begin{align}
 \text{(B)} &= \frac{1}{2}\,f_{[a}{}^{efg}\,\MA_{b]ep}\,\varepsilon_{fgqr}\,\MA^r\,\ell^a\wedge\ell^p\wedge\ell^q
 = -\frac{1}{3!}\,f_{[a}{}^{efg}\,\MA_{b]pq}\,\MA_{efg}\,\ell^a\wedge\ell^p\wedge\ell^q
\nn\\
 &= -\frac{1}{2\cdot 3!}\,f_a{}^{efg}\,\varepsilon_{bpqr}\,\MA^r\,\MA_{efg}\,\ell^a\wedge\ell^p\wedge\ell^q
  +\frac{1}{2\cdot 3!}\,f_b{}^{efg}\,\MA_{apq}\,\MA_{efg}\,\ell^a\wedge\ell^p\wedge\ell^q
\nn\\
 &= -\frac{1}{3!}\,\MA^r\,f_r{}^{efg}\,\MA_{efg}\,\frac{1}{3!}\,\varepsilon_{bapq}\,\ell^a\wedge\ell^p\wedge\ell^q
  + \frac{1}{3\cdot 3!}\,f_b{}^{efg}\,\MA_{apq}\,\MA_{efg}\,\ell^a\wedge\ell^p\wedge\ell^q
\nn\\
 &= \frac{1}{3}\,f_b{}^{efg}\,\MA_{efg}\,\MA_3 - \frac{1}{3!}\,n_b\, \MA^r\,f_r{}^{efg}\,\MA_{efg} \,,
\end{align}
where we have defined $n_a \equiv \frac{1}{3!}\,\varepsilon_{abcd}\,\ell^b\wedge\ell^c\wedge\ell^d$, which satisfies $H_a^b\,n_b = 0$\,. 
The third term is rather complicated and it may not be useful to show all of the computation. 
By using the Schouten identity, we find
\begin{align}
 \text{(C)} = \Bigl(\frac{1}{3}\,f_b{}^{efg}\,\MA_{efg} - f_b{}^{efg}\,\MA_{fgp}\,H^p_e \Bigr) *1 
 + N_b^a\,\Bigl(\frac{1}{2}\,f_e{}^{efg}\,\MA_{afg} - \frac{1}{3}\,f_a{}^{efg}\, \MA_{efg} \Bigr) *1 \,.
\end{align}
Then, by acting the projection, which removes terms proportional to $n_a$ and $N_a^b$, we obtain
\begin{align}
 &H_a^c\,\ell^b\wedge \Bigl(\frac{1}{2}\,f_{[b}{}^{def}\,P_{c]d}\wedge P_{ef}\Bigr)
\nn\\
 &= - \frac{1}{3}\,H_a^c\,f_c{}^{efg}\,\Bigl[\ell_e\wedge \ell_f\wedge \ell_g 
 + \bigl(3\,\MA_{fgp}\,H^p_e - \MA_{efg}\bigr) *1
 - \MA_{efg}\,\MA_3 \Bigr] \,.
\end{align}
This is precisely the desired equation \eqref{eq:id-n=4}.

\end{document}